\documentclass[lettersize,journal]{IEEEtran}
\usepackage{times}
\usepackage{graphicx}
\usepackage{amsmath, amsfonts}
\usepackage{amssymb}
\usepackage{algorithm}
\usepackage{algpseudocode}
\usepackage{multirow}
\usepackage{bm}
\usepackage{float}
\usepackage{lettrine}
\usepackage{stfloats}
\usepackage{cite}
\usepackage{amsmath}
\usepackage{diagbox}
\usepackage{makecell}
\usepackage{subfigure}
\usepackage{caption}
\usepackage{subfig} 
\usepackage{xcolor}
\usepackage{url}
\usepackage{nomencl}
\usepackage{colortbl}
\makenomenclature

\usepackage{enumitem}
\usepackage{booktabs}
\usepackage{tabularx}
\usepackage{xspace}

\newcommand{\eg}{\emph{e.g.,}\xspace}

\begin{document}

\title{Large Language Model Enhanced Graph Invariant Contrastive Learning for Out-of-Distribution Recommendation}

\author{Jiahao Liang\IEEEauthorrefmark{2}, Haoran~Yang\IEEEauthorrefmark{2}, Xiangyu~Zhao, \IEEEmembership{Member,~IEEE},
Zhiwen~Yu, \IEEEmembership{Senior Member,~IEEE},\\
% C. L. Philip Chen, \IEEEmembership{Fellow,~IEEE}, 
Mianjie Li, Chuan Shi, \IEEEmembership{Senior Member,~IEEE},
Kaixiang Yang\IEEEauthorrefmark{1}, \IEEEmembership{Member,~IEEE}
\IEEEcompsocitemizethanks{
% \IEEEcompsocthanksitem Kaixiang Yang is with the State Key
% Laboratory of Industrial Control Technology,  College of Control Science and Engineering in Zhejiang University, China. Email: yangkaixiang@zju.edu.cn.
% \IEEEcompsocthanksitem Zhiwen Yu, Yuchen Liu and C. L. Philip Chen are with the School of Computer Science and Engineering in South China University of Technology, China.  Email: zhwyu@scut.edu.cn.(Corresponding author: Zhiwen Yu.)
\IEEEcompsocthanksitem Jiahao Liang, Kaixiang Yang are with the School of Computer Science and Engineering in South China University of Technology, China. (Corresponding author: Kaixiang Yang, e-mail: csjiahliang6@mail.scut.edu.cn, yangkx@scut.edu.cn)
\IEEEcompsocthanksitem Haoran Yang is with the School of Computer Science and Engineering at Central South University, Changsha, China. (e-mail: yhr.cse@csu.edu.cn)
\IEEEcompsocthanksitem Xiangyu Zhao is with City University of Hong Kong, Department of Data Science, Hong Kong. (Corresponding author: Xiangyu Zhao, e-mail: xy.zhao@cityu.edu.hk.)
\IEEEcompsocthanksitem Zhiwen Yu is with the School of Computer Science and Engineering in South China University of Technology and the Pengcheng Lab, China. ( e-mail:zhwyu@scut.edu.cn)
\IEEEcompsocthanksitem Mianjie Li is with the School of Electronics and Information, Guangdong Polytechnic Normal University, Guangzhou, 510665, China. (e-mail: mianjieli@gpnu.edu.cn)
% \IEEEcompsocthanksitem Xiaoqing Liu is with the School of Future Technology in South China University of Technology, Guangzhou 510641, and also with Pengcheng Laboratory, Shenzhen 518000, China (e-mail: ft\_liuxiaoqing@mail.scut.edu.cn).
% \IEEEcompsocthanksitem C.L.Philip Chen is with the School of Computer Science and Engineering in South China University of Technology and the Pazhou Lab, China (e-mail: philip.chen@ieee.org.)
\IEEEcompsocthanksitem Chuan Shi is with School of Computer Science, Beijing University of Posts and Telecommunications, Beijing, China (e-mail: shichuan@bupt.edu.cn.)
}

% <-this % stops a space
\thanks{\IEEEauthorrefmark{1} indicates the corresponding authors. \IEEEauthorrefmark{2} indicates the equal contribution.}}

\markboth{IEEE Transactions on Knowledge and Data Engineering}
{Shell \MakeLowercase{\textit{et al.}}: IEEE Transactions on Knowledge and Data Engineering}

\IEEEcompsoctitleabstractindextext{

\begin{abstract}
%\yhr{(i) Improve the overall typesetting, especially pay attention to the end of each paragraph, no less than 4 words. (ii) Improve the writing quality by inputting every paragraph to ChatGPT for refining. These two things should be done when you complete all the other revisions.}
Out-of-distribution (OOD) generalization has emerged as a significant challenge in graph recommender systems. Traditional graph neural network algorithms often fail because they learn spurious environmental correlations instead of stable causal relationships, leading to substantial performance degradation under distribution shifts. While recent advancements in Large Language Models (LLMs) offer a promising avenue due to their vast world knowledge and reasoning capabilities, effectively integrating this knowledge with the fine-grained topology of specific graphs to solve the OOD problem remains a significant challenge. To address these issues, we propose {$\textbf{Inv}$ariant $\textbf{G}$raph $\textbf{C}$ontrastive Learning with $\textbf{LLM}$s for Out-of-Distribution Recommendation (InvGCLLM)}, an innovative causal learning framework that synergistically integrates the strengths of data-driven models and knowledge-driven LLMs. Our framework first employs a data-driven invariant learning model to generate causal confidence scores for each user-item interaction. These scores then guide an LLM to perform targeted graph refinement, leveraging its world knowledge to prune spurious connections and augment missing causal links. Finally, the structurally purified graphs provide robust supervision for a causality-guided contrastive learning objective, enabling the model to learn representations that are resilient to spurious correlations. Experiments conducted on four public datasets demonstrate that InvGCLLM achieves significant improvements in out-of-distribution recommendation, consistently outperforming state-of-the-art baselines.
\end{abstract}

}
\maketitle
\IEEEdisplaynotcompsoctitleabstractindextext

\IEEEpeerreviewmaketitle

%\clearpage
%\appendix
%\input{6Appendix}

\section{Introduction}
In today's rapidly evolving field of Recommender System (RS), GNNs have gained a great deal of attention ~\cite{wang2023diffusion,wang2019kgat,yang2024generate,gao2023autotransfer,WANG2025113415,yuan2025hyperbolic}, because of their ability to simulate complex interactions between various entities in real-world scenarios~\cite{jin2022empowering,wang2022adaptive,zhao2024masked}. Specifically, GNNs efficiently processes graphs of different sizes and learns low-dimensional embeddings by iteratively propagating and aggregating the semantic information of topological neighbors~\cite{wu2022graph}. While many graph recommendation algorithms perform well on In-Distribution (ID) data~\cite{xu2022category,fan2022graph,yang2024empirical,zhang2024linear}, maintaining stable performance in the face of OOD data remains a major challenge for the rapid development of RS.

\begin{figure}[h]
\centering
\includegraphics[width=\linewidth]{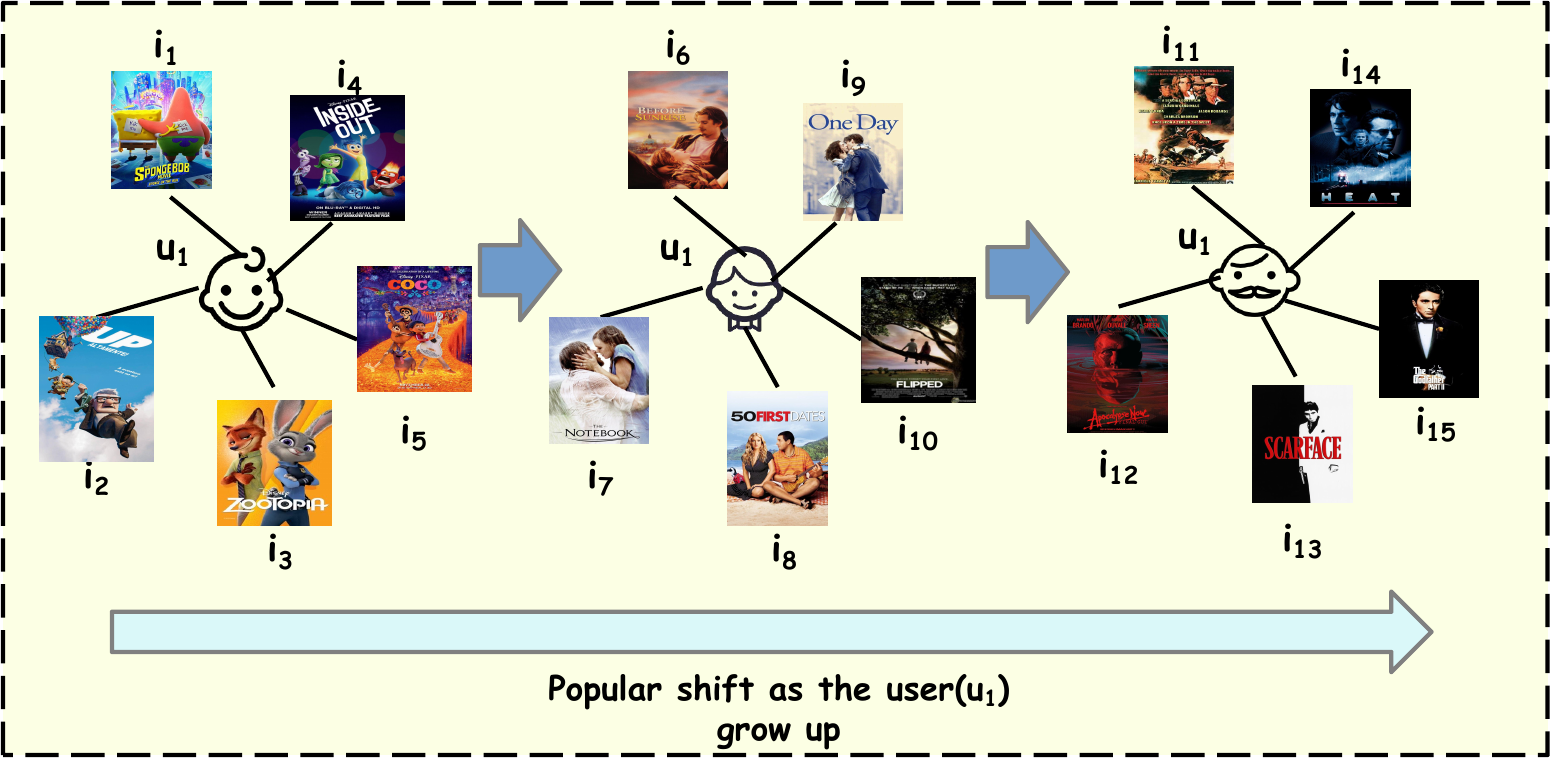}
\caption{Illustration of user preference shift over time, highlighting the OOD recommendation challenge as $u_1$'s favored movie genres change with age.}
\label{fig:Fig_OOD_Rec}
\end{figure}

In the context of RS, the OOD recommendation problem refers to the challenge of accurately predicting user preferences when the data distribution shifts over time or across domains~\cite{yang2025dual,zhao2025graph}. As illustrated in Figure~\ref{fig:Fig_OOD_Rec}, the user-item interaction graph for a user $u_1$ evolves as the user grows up. For example, during childhood, $u_1$ primarily interacts with cartoon movies such as \textit{Up} ($i_2$), \textit{Zootopia} ($i_3$), \textit{Coco} ($i_5$), and \textit{Inside Out} ($i_4$). As $u_1$ enters adolescence, his preferences shift toward romantic movies like \textit{The Notebook} ($i_7$), \textit{One Day} ($i_9$), \textit{Flipped} ($i_{10}$), and \textit{50 First Dates} ($i_8$). In adulthood, $u_1$ begins to favor genres such as westerns and gangster films, including classics starring Al Pacino, such as \textit{Scarface} ($i_{13}$), as well as movies like \textit{The Godfather} ($i_{15}$), \textit{Heat} ($i_{14}$), and \textit{Apocalypse Now} ($i_{12}$). This dynamic demonstrates a distributional shift in user-item interactions, where the system must learn representations that can generalize across different stages of user preferences. The core challenge is to ensure that the recommendation model does not overfit to static or historical preferences, but instead adapts to evolving user interests, which is crucial for OOD recommendation~\cite{yang2025dual,zhao2025graph,zhang2024disentangled,zhao2025distributionally}.

The root of the OOD problem lies in the tendency of models to learn spurious correlations present in the training environment, rather than the stable, underlying causal relationships that govern user preferences. For instance, a model might learn a spurious association that 'users in their teens prefer romantic movies', but fail to capture the invariant causal factor, such as a preference for 'compelling storytelling' which remains stable even as the user's taste in genres evolves from cartoons in childhood to dramas in adulthood. As shown in Figure~\ref{fig:Fig_OOD_Rec}, a user's interests shift across different life stages. A model that overfits to the correlations of one stage will inevitably fail in another. Therefore, the key to solving the graph OOD problem is to learn representations that are insensitive to distribution shifts by explicitly disentangling causal factors from non-causal, spurious associations.

Existing methods, however, fall short of disentangling causal factors from spurious associations, primarily due to a lack of explicit supervision for causal separation in the representation space~\cite{zhang2024disentangled,zhao2025distributionally,LIANG2025103792}. Without direct signals to align representations with causal factors while repelling non-causal ones, models often learn superficial invariants.  To address this supervision deficit, Causal Contrastive Learning (CCL) has emerged as a promising framework that explicitly guides representation learning by constructing causal positive and non-causal negative samples~\cite{10.1609/aaai.v38i8.28738,11128741,kim2025causality}. The efficacy of CCL, however, is entirely contingent on the causal purity of its contrastive views. While current CCL methods surpass the indiscriminate perturbations of traditional Graph Contrastive Learning (GCL)~\cite{wu2021self,yu2022graph,cai2023lightgcl}, they remain fundamentally limited by their reliance on purely data-driven signals. Confined to the statistical patterns within the training data, they lack an external frame of reference to distinguish deeply entrenched spurious correlations from genuine causal links. For instance, a model might learn that user interaction with 'blockbuster films' is an invariant feature, when this is merely a stable proxy for a true causal preference, such as 'an affinity for high production value cinematography'. Such a model is brittle, as its performance collapses when this proxy correlation weakens in a new environment. This vulnerability can lead to the generation of contaminated causal views, ultimately undermining the model's ability to achieve robust Out-of-Distribution (OOD) generalization.

A natural solution is to introduce Large Language Models (LLMs) for intelligent graph structure editing, leveraging their rich world knowledge. However, the direct application of LLMs presents its own challenges. Despite possessing vast general knowledge, LLMs lack fine-grained insight into the specific structural patterns and user behavior dynamics of a given dataset. Without effective data-driven guidance, their edits risk being arbitrary or even contradictory to the data's intrinsic regularities, thus failing to reliably improve the graph's causal purity.

% However, the success of this framework is entirely dependent on the quality of its contrastive views, which exposes a fundamental weakness in mainstream Graph Contrastive Learning (GCL) methods: their data augmentation process is blind~\cite{wu2021self,yu2022graph,cai2023lightgcl}. Traditional GCL relies on random structural perturbations, such as edge dropping, to create views. This approach cannot discern the causal nature of interactions and is therefore incapable of generating the semantically pure contrastive samples required to serve OOD generalization objectives.

To overcome these challenges, we move beyond incremental improvements and propose an innovative causal learning framework that synergistically integrates the strengths of data-driven models and knowledge-driven LLMs. Our method, \textbf{Inv}ariant \textbf{G}raph \textbf{C}ontrastive Learning with \textbf{LLM}s for Out-of-Distribution Recommendation, namely \textbf{InvGCLLM}, pioneers a structured, multi-stage process. The first stage is data-driven causal score generation, where we employ an invariant learning model to efficiently analyze the user-item graph and assign a causal confidence score to each edge, producing high-quality initial proposals. These scores then guide our novel \textbf{LLM-Calibrated Causal Graph Editing (LLMC2GE)} module. Here, an LLM leverages its embedded real-world knowledge to verify and calibrate the relationships identified by the scores, making informed decisions to prune spurious edges and augment missing causal links. Finally, these LLM-refined, structurally purified causal and non-causal graph views serve as the foundation for our \textbf{Causal-Informed Contrastive Learning (CICL)} objective. This contrastive loss explicitly guides the model to learn representations that align with causal patterns while discriminating against spurious ones, thereby fostering robust Out-of-Distribution (OOD) generalization.

To verify the validity of our approach, we designed additional experiments using traditional graph contrast learning to demonstrate the poor performance of traditional methods on OOD problems. Our experimental results show that the LLM-enhanced contrast learning algorithm outperforms all baselines in OOD problems and significantly improves the generalization ability of the recommendation in the OOD setting.  Our contribution is multifaceted:

\begin{itemize}[leftmargin=*]
\item We introduce InvGCLLM, a novel causal learning framework that, for the first time, effectively synergizes data-driven causal contrastive learning with the world knowledge of Large Language Models (LLMs) to achieve robust causal learning on graphs.
\item We design a specific LLM-based graph refinement mechanism (LLMC2GE)). In this mechanism, the LLM purifies the graph structure, with its edits being verified and calibrated by data-driven causal scores. This process ensures the refined graph guides a more targeted and effective contrastive learning objective.
\item Extensive experiments on several benchmark datasets demonstrate that InvGCLLM significantly outperforms state-of-the-art methods in Out-of-Distribution (OOD) recommendation scenarios.
\end{itemize}

\section{Problem Formulation}

\subsection{Preliminaries}
Let $\mathcal{G} = (\mathcal{V}, \mathcal{A}, \mathbf{X})$ represent a user-item bipartite graph, where $\mathcal{V} = \mathcal{V}_U \cup \mathcal{V}_I$ is the set of user and item nodes. $\mathcal{A} \in \{0, 1\}^{|\mathcal{V}| \times |\mathcal{V}|}$ is the adjacency matrix representing interactions, and $\mathbf{X}$ denotes the matrix of node features.

A graph neural network (GNN), denoted as $f_\theta$, serves as a node encoder that maps the graph structure and node features to a low-dimensional embedding space:
\begin{equation}
    \mathbf{Z} = f_\theta(\mathcal{A}, \mathbf{X}),
\end{equation}
where $\mathbf{Z} \in \mathbb{R}^{|\mathcal{V}| \times d}$ is the resulting node embedding matrix. The function $f_\theta$ is pre-trained on a source (training) graph, $\mathcal{G}_{tr} = (\mathcal{V}_{tr}, \mathcal{A}_{tr}, \mathbf{X}_{tr})$, by optimizing a standard recommendation loss function, $\mathcal{L}_{rec}$ (\eg BPR loss), over its parameters $\theta$:
\begin{equation}
    \theta^* = \arg\min_{\theta} \mathcal{L}_{rec}(f_\theta(\mathcal{A}_{tr}, \mathbf{X}_{tr})).
\end{equation}

\subsection{Out-of-Distribution Graph Recommendation Task}
We address the task of out-of-distribution (OOD) graph recommendation at test time. We are given the pre-trained GNN encoder $f_{\theta^*}$ and an unlabeled target (test) graph $\mathcal{G}_{te} = (\mathcal{V}_{te}, \mathcal{A}_{te}, \mathbf{X}_{te})$.

The core assumption is that the training and test graphs are drawn from different distributions. Formally, the joint probability distribution of the graph structure and features in the target domain differs from that of the source domain:
\begin{equation}
    P_{te}(\mathcal{A}, \mathbf{X}) \neq P_{tr}(\mathcal{A}, \mathbf{X}).
\end{equation}
This distribution shift means that the pre-trained model $f_{\theta^*}$ is no longer optimal for the target graph $\mathcal{G}_{te}$.

The objective is to devise a test-time adaptation strategy that can generate high-quality user and item embeddings for the target graph, $\mathbf{Z}_{te}$, to be used for final recommendations. This must be accomplished without access to any ground-truth labels or performing gradient-based training on the test graph $\mathcal{G}_{te}$. The performance of the adaptation is evaluated using standard recommendation metrics (\eg NDCG, Recall) on the held-out interactions in $\mathcal{G}_{te}$.

\section{Framework}
\begin{figure*}[t]
  \centering
  \includegraphics[width=\textwidth]{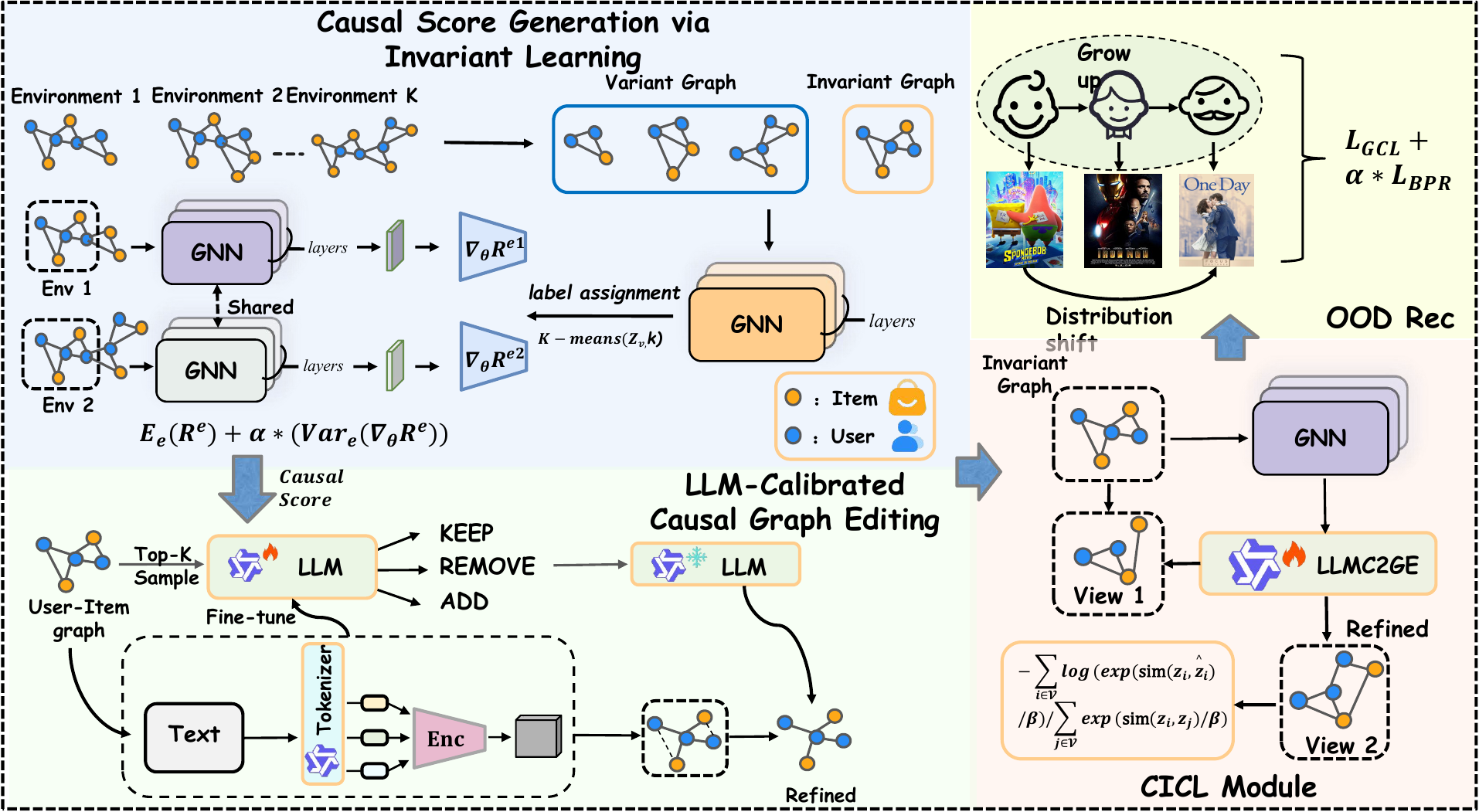}
  \caption{The InvGCLLM framework: Environment Extractor identifies invariant and variant features, Invariant Learning Module optimizes stable representations across environments, LLMGCL Module generates contrastive views, and Recommendation Module delivers robust out-of-distribution (OOD) recommendations.}
  \label{fig:Fig2_overall}
\end{figure*}

In this work, we propose InvGCLLM, a contrastive learning algorithm augmented by large language models (LLMs) and grounded in invariance principles to address out-of-distribution (OOD) challenges in recommender systems, as shown in Figure~\ref{fig:Fig2_overall}. InvGCLLM consists of three main components: (1) an Environment Extractor that distinguishes user and item features across different distribution environments; (2) an Invariant Learning Module that enhances generalization by learning invariant representations from graph data; and (3) the LLMGCL Module, where our Causal-aware LLM Structure Enhancement (LLMC2GE) leverages LLMs to edit graph structures, enabling the model to capture stable causal relationships and mitigate the effects of distribution shifts on downstream tasks. Through the integration of these modules, InvGCLLM improves recommendation robustness and adaptability across diverse environments, significantly enhancing system performance and stability.

\subsection{Causal Score Generation via Invariant Learning}
To provide a robust, data-driven foundation for our subsequent LLM-based graph refinement, we first introduce a module dedicated to generating a causal score for each user-item interaction. This score, denoted by the matrix $\mathbf{S}$, quantifies the stability of an interaction across different underlying contexts or ‘environments.‘ The core idea is that truly causal relationships should remain invariant despite distribution shifts, whereas spurious correlations will not. This module achieves its goal through a two-stage process. First, the \textbf{Environment Extractor} identifies latent environments within the data based on interaction patterns. Second, the \textbf{Invariant Learning Module} leverages these identified environments to train a generator that assigns a high causal score to those interactions that prove stable across all environments.

\subsubsection{Environment Extractor}
To address the challenge that traditional GCL methods cannot distinguish environment labels, leading to difficulties in effectively handling distribution shifts, we propose the Environment Extractor. In recommender systems, user-item interactions are modeled as a graph $\mathcal{G} = (\mathcal{V}, \mathcal{E})$. The purpose of the Environment Extractor is to identify hidden environmental factors that affect user decisions, thereby improving the model's robustness.

We assume that within the graph $\mathcal{G}$, there exists a set of invariant information $\mathcal{M}_I$ that consistently reflects user preferences. To identify this, we introduce a learnable generator $\Phi(\mathcal{G})$, which decomposes the graph into invariant information $\mathcal{M}_I$ and variant information $\mathcal{M}_V$:
\begin{equation}
\label{eq:mask_decomposition}
\mathcal{M}_I = \text{Top}\tau(\mathbf{M_s} \odot \mathbf{M_g}), \quad \mathcal{M}_V = \mathbf{M_g} - \mathcal{M}_I
\end{equation}
where $\mathbf{M_g}$ is the adjacency matrix of the original graph and $\odot$ denotes element-wise multiplication. The soft mask matrix $\mathbf{M_s}$ contains learned probabilities indicating the invariance of each interaction, and the $\text{Top}\tau(\cdot)$ operation selects the top-$\tau$ most stable interactions.

The generation of the soft mask $\mathbf{M_s}$ is a critical learnable process. The soft mask $\mathbf{M_s}$ is generated by a learnable module. First, we employ a masking GNN, denoted as $\text{GNN}_{\text{mask}}$, to compute structural embeddings for all nodes in the input graph $\mathcal{G}$. An invariance score for each user-item pair $(u,i)$ is then derived by feeding the concatenation of their embeddings into a small Multi-Layer Perceptron (MLP) followed by a Sigmoid activation function:
\begin{equation}
    s_{u,i} = \sigma(\text{MLP}(\mathbf{h}_{\text{mask},u} \oplus \mathbf{h}_{\text{mask},i})),
\end{equation}
where $s_{u,i}$ is the corresponding element in $\mathbf{M_s}$. The parameters of $\text{GNN}_{\text{mask}}$ and the MLP are optimized end-to-end with the main learning objective, which enables the model to dynamically identify invariant interactions.

We infer the underlying environments from the variant information $\mathcal{M}_V$ by first generating embeddings for the variant interactions and then clustering them to derive environment labels.

\paragraph{Step 1: Variant Interaction Embedding Generation} First, we obtain a vector representation for each interaction in the variant subgraph. We employ a GNN encoder, denoted as $\text{GNN}_{\text{enc}}$, to compute node embeddings based on the variant graph structure, yielding $\mathbf{H}_V = \text{GNN}_{\text{enc}}(\mathcal{M}_V, \mathbf{X})$, where $\mathbf{X}$ represents initial node features. For each interaction $(u, i)$ in the variant edge set $\mathcal{E}_V$, we construct its interaction embedding $\mathbf{z}_{u,i}$ by concatenating the corresponding node embeddings from $\mathbf{H}_V$. This results in a set of interaction embeddings $\mathbf{Z}_V = \{ \mathbf{z}_{u,i} | (u,i) \in \mathcal{E}_V \}$.

\paragraph{Step 2: Clustering and Environment Partitioning}
Next, we partition these interaction embeddings using the standard K-means clustering algorithm~\cite{hartigan1979algorithm}. The algorithm takes the set of embeddings $\mathbf{Z}_V$ and a hyperparameter for the number of environments, $K$, as input. It directly outputs the final environment partition $\mathcal{E}_{\text{infer}}$ as follows:
\begin{equation}
    \mathcal{E}_{\text{infer}} = \text{K-means}(\mathbf{Z}_V, K).
\end{equation}
The output, $\mathcal{E}_{\text{infer}} = \{e_1, e_2, \dots, e_K\}$, is a partition of the variant interactions, where each set $e_k$ contains the interactions belonging to the $k$-th cluster. This partition serves as the environment labels required by the subsequent module.

\subsubsection{Invariant Learning Module}
After extracting the invariant subgraphs and inferring environment labels $\mathcal{E}_{\text{infer}}$, the Invariant Learning Module aims to enhance generalization by learning user preferences that are invariant across the varying environments. The inferred partition $\mathcal{E}_{\text{infer}}$ is critical for this stage, as the expectation $E_e(R^e)$ and variance $\text{Var}_e(\nabla_\theta R^e)$ terms in our objective are computed across these $K$ data-driven environments.

The goal is to learn an optimal generator $\Phi^*$ that ensures consistency across different environments. Formally, we define the set of invariant generators $\mathcal{I}$ as satisfying the condition $P^{e}(Y | \Phi(\mathcal{M})) = P^{e'}(Y | \Phi(\mathcal{M}))$ for any two environments $e$ and $e'$. We find the optimal generator $\Phi^*$ by minimizing the discrepancy between environments:
\begin{equation}
\label{eq:optimal_mapping}
\Phi^* = \arg \min_{\Phi \in \mathcal{I}_{\mathcal{E}}} E_e(R^e) + \alpha \cdot \text{Var}_e(\nabla_\theta R^e),
\end{equation}
where $R^e$ is the risk function in environment $e$, and the variance term encourages stability of the learned representations.

To optimize the generator $\Phi$, which produces the causal score matrix $\mathbf{S}$, we minimize an invariant learning objective. The score matrix $\mathbf{S}$ reflects the stability of interactions, with higher scores indicating greater invariance. The objective is:
\begin{equation}
\label{eq:objective_function}
E_e(R^e) + \alpha \cdot \text{Var}_e(\nabla_\theta R^e) + \beta_e \cdot \mathcal{L}_{\text{score}},
\end{equation}
where $\alpha$ is a balancing hyperparameter and $\mathcal{L}_{\text{score}}$ is a score regularization loss:
\begin{align}
\label{eq:score_loss}
\mathcal{L}_{\text{score}} = -\frac{1}{|\mathcal{E}|} \sum_{(u,i) \in \mathcal{E}} \bigg[ & S_{u,i} \log P^{e}(Y_{u,i} | \mathcal{M}_I) \notag \\
& + (1 - S_{u,i}) \log (1 - P^{e}(Y_{u,i} | \mathcal{M}_I)) \bigg].
\end{align}
This loss ensures that higher scores $S_{u,i}$ are assigned to interactions that are stable across environments. By optimizing this objective, we obtain the desired generator $\Phi$, which produces both the invariant information $\mathcal{M}_I$ and the final causal score matrix $\mathbf{S}$.

\subsection{LLM-Calibrated Causal Graph Editing}
To overcome the limitations of purely data-driven heuristics in the causal score matrix $\mathbf{S}$, we introduce the LLM-Calibrated Causal Graph Editing (LLMC2GE) module. This module leverages a Large Language Model (LLM) to verify and calibrate the initial causal scores, infusing statistical evidence with its embedded world knowledge. This ensures the resulting graph views are grounded in both quantitative analysis and rich semantic judgment, significantly enhancing their causal purity.

The refinement process begins by using the causal score matrix $\mathbf{S}$ to select candidate edges from the original graph $\mathcal{G} = (\mathcal{V}, \mathcal{E})$ for LLM adjudication. Formally, we define two candidate sets: a set of potential causal links, $\mathcal{E}^+_c$, and a set of potential spurious links, $\mathcal{E}^-_c$:
\begin{align}
\mathcal{E}^+_c &= \{ (u,i) \in \mathcal{E} \mid S_{ui} \in \text{Top-K}(\mathbf{S}) \}, \\
\mathcal{E}^-_c &= \{ (u,i) \in \mathcal{E} \mid S_{ui} \in \text{Bottom-K}(\mathbf{S}) \}.
\end{align}
This step efficiently focuses the LLM's reasoning capabilities on the most impactful edges.

Subsequently, for each candidate edge $(u,i)$ in these sets, we prompt the LLM to perform a semantic evaluation. We construct a structured prompt $P(u,i)$ containing the textual attributes of the nodes ($T_u, T_i$) and the data-driven causal score $S_{ui}$. The LLM then acts as an adjudication function, $f_{\text{LLM}}$, which processes the prompt and returns a discrete decision $d_{ui} \in \{\text{KEEP, REMOVE}\}$. For high-score candidates in $\mathcal{E}^+_c$, the LLM is asked to verify a stable causal link, while for low-score candidates in $\mathcal{E}^-_c$, it is asked to confirm a spurious correlation. Based on the returned decisions, we define the set of confirmed spurious edges to be removed:
\begin{equation}
\mathcal{E}_{\text{remove}} = \{ (u,i) \in \mathcal{E}^-_c \mid f_{\text{LLM}}(P(u,i)) = \text{REMOVE} \}.
\end{equation}
Optionally, the LLM can also be prompted to suggest new, causally-consistent links for high-confidence nodes, generating a set of edges to be added, $\mathcal{E}_{\text{add}}$.

Finally, based on the LLM's adjudications, we construct two topologically refined graph views for the subsequent contrastive learning stage. The \textbf{Causal View}, $\mathcal{G}_c = (\mathcal{V}, \mathcal{E}_c)$, is constructed by pruning the confirmed spurious edges from the original graph and augmenting it with newly proposed causal links:
\begin{equation}
\mathcal{E}_c = (\mathcal{E} \setminus \mathcal{E}_{\text{remove}}) \cup \mathcal{E}_{\text{add}}.
\end{equation}
This view represents our best estimate of the pure causal skeleton. Concurrently, the \textbf{Spurious View}, $\mathcal{G}_s = (\mathcal{V}, \mathcal{E}_{\text{remove}})$, is composed exclusively of the edges confirmed by the LLM as spurious, providing a concentrated set of negative signals for debiasing.

\subsection{Causal-Informed Contrastive Learning}
Having obtained high-purity causal ($\mathcal{G}_c$) and spurious ($\mathcal{G}_s$) views via LLM-based editing, we introduce a contrastive learning objective specifically designed for OOD generalization. Our key motivation is to ‘purify‘ the main representation $\mathbf{h}_v$ (learned from the original, noisy graph $\mathcal{G}$) by explicitly teaching the model to distinguish between causal and non-causal factors.

\paragraph{Multi-View Representation Encoding.}
To implement this, we employ a single GNN encoder, $f_{\text{GNN}}(\cdot; \Theta)$, which is shared across all graph views to ensure a consistent embedding space. In each training batch, the encoder processes the original node features $\mathbf{X}$ using three different graph structures to generate three distinct representations for each node $v$: the \textbf{main representation ($\mathbf{h}_v$)} from the original graph $\mathcal{G}$, the \textbf{causal representation ($\mathbf{h}_{v,c}$)} from the causal view $\mathcal{G}_c$, and the \textbf{spurious representation ($\mathbf{h}_{v,s}$)} from the spurious view $\mathcal{G}_s$. Formally:
\begin{align}
\mathbf{h}_v &= f_{\text{GNN}}(\mathcal{G}, \mathbf{X}; \Theta), \\
\mathbf{h}_{v,c} &= f_{\text{GNN}}(\mathcal{G}_c, \mathbf{X}; \Theta), \\
\mathbf{h}_{v,s} &= f_{\text{GNN}}(\mathcal{G}_s, \mathbf{X}; \Theta).
\end{align}

\paragraph{Causal-Informed Contrastive Objective.}
We design a Causal-Informed Contrastive Loss ($\mathcal{L}_{\text{CICL}}$) to purify the main representation $\mathbf{h}_u$. For a given anchor node $u$, the objective maximizes its agreement with its causal counterpart while minimizing agreement with a set of negative samples.
The \textbf{positive pair} is ($\mathbf{h}_u, \mathbf{h}_{u,c}$), which encourages the main representation to absorb stable, causal information. The \textbf{negative set} includes standard in-batch negatives ($\{\mathbf{h}_v\}_{v \neq u}$) and, critically, a \textbf{self-negative} pair ($\mathbf{h}_u, \mathbf{h}_{u,s}$). Pushing the main representation away from its own spurious version explicitly forces the model to disentangle and discard spurious patterns.

This contrastive scheme is implemented by minimizing the loss function defined below. The denominator term $Z_u$ sums the scores of the positive pair and all negative pairs:
\begin{align}
\label{eq:norm_term}
Z_u = \underbrace{\exp\!\left(\frac{\text{sim}(\mathbf{h}_u, \mathbf{h}_{u,c})}{\tau}\right)}_{\text{Positive Pair}}
& + \underbrace{\sum_{v \in \mathcal{B}, v \ne u}
\exp\!\left(\frac{\text{sim}(\mathbf{h}_u, \mathbf{h}_v)}{\tau}\right)}_{\text{In-batch Negatives}} \notag \\
& + \underbrace{\exp\!\left(\frac{\text{sim}(\mathbf{h}_u, \mathbf{h}_{u,s})}{\tau}\right)}_{\text{Self-Negative}}
\end{align}
where $\mathcal{B}$ is the batch of nodes, $\text{sim}(\cdot, \cdot)$ is cosine similarity, and $\tau$ is a temperature parameter. The final contrastive loss over the batch is:
\begin{equation}
\label{eq:cicl_loss_final}
\mathcal{L}_{\text{CICL}} = -\sum_{u \in \mathcal{B}} \log \frac{\exp(\text{sim}(\mathbf{h}_u, \mathbf{h}_{u,c}) / \tau)}{Z_u}
\end{equation}
Through minimizing this objective, the main representations $\{\mathbf{h}_v\}$ are purified. These purified representations are used exclusively for the final recommendation task.

\subsection{Recommendation and Optimization}

The final stage of our framework uses the purified embeddings for the downstream recommendation task. The overall model is trained end-to-end by optimizing a total objective function that combines our proposed Causal-Informed Contrastive Loss with a standard recommendation loss.

The final learning objective is a weighted sum of the contrastive loss $\mathcal{L}_{\text{CICL}}$ and the Bayesian Personalized Ranking (BPR) loss $\mathcal{L}_{\text{BPR}}$, which optimizes the ranking of items for users:
\begin{equation}
\label{eq:total_loss}
\mathcal{L}_{\text{total}} = \mathcal{L}_{\text{CICL}} + \lambda \mathcal{L}_{\text{BPR}},
\end{equation}
where $\lambda$ is a hyperparameter balancing the two objectives. The BPR loss is defined as:
\begin{equation}
\label{eq:bpr_loss}
\mathcal{L}_{\text{BPR}} = -\sum_{(u,i,j) \in \mathcal{D}} \ln \sigma(\hat{y}_{ui} - \hat{y}_{uj}),
\end{equation}
where $\mathcal{D}$ is the set of training triplets $(u, i, j)$, such that user $u$ has interacted with item $i$ (a positive sample) but not item $j$ (a negative sample). The prediction score $\hat{y}_{ui}$ is calculated using the dot product of the purified user and item embeddings learned from the CICL stage: $\hat{y}_{ui} = \mathbf{h}_u^\top \mathbf{h}_i$. The function $\sigma(\cdot)$ is the sigmoid function. By minimizing $\mathcal{L}_{\text{total}}$, the model learns representations that are not only debiased and robust to distribution shifts but also optimized for producing accurate personalized rankings.

\section{Theoretical Analysis}\label{sec:theory}

The proposed framework combines the \textbf{Environment Extractor} and \textbf{Invariant Learning Module} to effectively distinguish between variant and invariant patterns in user-item interaction graphs for out-of-distribution (OOD) recommender systems. This section provides a theoretical analysis of how these modules synergistically achieve this goal, leveraging graph decomposition, environment inference, and invariant optimization. We formalize the processes with mathematical rigor, introducing additional equations to elucidate the mechanisms behind pattern discrimination.

\subsection{Environment Extractor: Decomposition and Environment Inference}

The \textbf{Environment Extractor} decomposes the interaction graph \(\mathcal{G} = (\mathcal{V}, \mathcal{E})\) into invariant (\(\mathcal{M}_I\)) and variant (\(\mathcal{M}_V\)) information, followed by environment inference to capture distributional contexts. This module ensures that stable user preferences are separated from environment-specific variations, enabling robust generalization.

\subsubsection{Decomposition of Interaction Patterns}

The decomposition process begins with the interaction matrix \(\mathbf{M} \in \mathbb{R}^{(|\mathcal{U}|+|\mathcal{I}|) \times (|\mathcal{U}|+|\mathcal{I}|)}\), where \(\mathcal{U}\) and \(\mathcal{I}\) represent users and items, respectively. The learnable generator \(\Phi(\mathcal{G})\) produces a soft mask matrix \(\mathbf{M_s}\), which is used to identify invariant patterns via:
\begin{equation}
\mathcal{M}_I = \text{Top}\tau(\mathbf{M_s} \odot \mathbf{M_g}), \quad \mathcal{M}_V = \mathbf{M_g} - \mathcal{M}_I,
\label{eq:mask_decomposition_theory}
\end{equation}
where \(\mathbf{M_g}\) is the adjacency matrix of \(\mathcal{G}\), \(\odot\) denotes element-wise multiplication, and \(\text{Top}\tau(\cdot)\) selects the top-\(\tau\) elements. The soft mask \(\mathbf{M_s}\) is generated by a graph neural network (GNN) as:
\begin{equation}
\mathbf{M_s} = \sigma(\text{GNN}_{\theta_s}(\mathbf{M_g}, \mathbf{X})),
\label{eq:soft_mask}
\end{equation}
where \(\sigma\) is a sigmoid activation, \(\mathbf{X} \in \mathbb{R}^{(|\mathcal{U}|+|\mathcal{I}|) \times d}\) is the node feature matrix, and \(\text{GNN}_{\theta_s}\) is parameterized by \(\theta_s\). The invariant information \(\mathcal{M}_I\) corresponds to edges with high mask values, representing stable interactions, while \(\mathcal{M}_V\) captures environment-sensitive interactions.

Theoretically, this decomposition is grounded in the assumption that invariant patterns reflect causal user preferences. To quantify the stability of \(\mathcal{M}_I\), we define the cross-environment consistency of an edge \(e_{u,i} \in \mathcal{E}\):
\begin{equation}
S(e_{u,i}) = \mathbb{E}_{e \sim \mathcal{E}} \left[ \frac{P^e(M_{u,i} = 1)}{P^e(M_{u,i} = 0)} \right],
\label{eq:edge_stability}
\end{equation}
where \(P^e(M_{u,i})\) is the probability of interaction under environment \(e\). Edges with high \(S(e_{u,i})\) are prioritized in \(\mathcal{M}_I\), as they are more likely to persist across distributions. The \(\text{Top}\tau\) operation ensures that \(\mathcal{M}_I\) includes only the most stable edges, minimizing the inclusion of spurious correlations.

The decomposition error is bounded by:
\begin{equation}
\|\mathbf{M_g} - (\mathcal{M}_I + \mathcal{M}_V)\|_F^2 = 0,
\label{eq:decomposition_error}
\end{equation}
where \(\|\cdot\|_F\) is the Frobenius norm, ensuring that the decomposition is lossless. However, the quality of separation depends on the GNN’s ability to learn \(\mathbf{M_s}\). We define the decomposition loss as:
\begin{equation}
\mathcal{L}_{\text{decomp}} = \mathbb{E}_{e \sim \mathcal{E}} \left[ S(e) \cdot M_{s,e} + (1 - S(e)) \cdot (1 - M_{s,e}) \right],
\label{eq:decomp_loss}
\end{equation}
which encourages high mask values for stable edges and low values for variant ones, aligning with the causal structure of the data.

\subsubsection{Environment Inference via Clustering}

The variant information \(\mathcal{M}_V\) is used to infer environmental contexts. First, a secondary GNN, parameterized by \(\theta_v\), generates embeddings \(\mathbf{H} \in \mathbb{R}^{N \times d}\) exclusively from this variant information:
\begin{equation}
\mathbf{H} = \text{GNN}_{\theta_v}(\mathcal{M}_V, \mathbf{X}),
\label{eq:variant_embedding}
\end{equation}
Subsequently, we apply standard K-means clustering to these embeddings. The algorithm partitions the embeddings into \(K\) clusters, representing \(K\) distinct environments, by minimizing the within-cluster variance, defined by the objective:
\begin{equation}
\min_{\mathbf{C}, \mathbf{\mu}} \sum_{k=1}^K \sum_{\mathbf{h}_i \in C_k} \|\mathbf{h}_i - \mathbf{\mu}_k\|_2^2,
\label{eq:kmeans_objective_theory}
\end{equation}
where \(C_k\) is the set of embeddings in the \(k\)-th cluster with centroid \(\mathbf{\mu}_k\).

The theoretical basis for this process relies on the assumption that \(\mathcal{M}_V\) successfully encodes environment-specific features. To formalize this, we can define the environment-specific divergence for an embedding \(\mathbf{h}_i\):
\begin{equation}
D(\mathbf{h}_i) = \mathbb{E}_{e, e'} \left[ \|\mathbf{h}_i^e - \mathbf{h}_i^{e'}\|_2^2 \right],
\label{eq:env_divergence}
\end{equation}
where \(\mathbf{h}_i^e\) is the embedding of an interaction under a true (but unobserved) environment \(e\). A high value for \(D(\mathbf{h}_i)\) indicates that the interaction's representation is sensitive to environmental changes, making it a strong signal for variant patterns. The K-means algorithm groups these high-divergence patterns into distinct clusters. The quality of this partitioning is assured as the clustering error is bounded:
\begin{equation}
\mathbb{E} \left[ \sum_{k=1}^K \sum_{\mathbf{h}_i \in C_k} \|\mathbf{h}_i - \mathbf{\mu}_k\|_2^2 \right] \leq \epsilon_{\text{cluster}},
\label{eq:cluster_error}
\end{equation}
where \(\epsilon_{\text{cluster}}\) is an error term dependent on the embedding quality and the number of clusters \(K\). This bound ensures that the variant patterns are partitioned accurately, enabling the model to reliably infer and capture distributional shifts from the data.

\subsection{Invariant Learning Module: Optimizing Stable Representations}

The \textbf{Invariant Learning Module} leverages the decomposed \(\mathcal{M}_I\) and inferred environment labels \(\mathcal{E}_{\text{infer}}\) to learn robust representations that generalize across environments. This module ensures that invariant patterns dominate the learned representations, while variant patterns are used to contextualize environment-specific adaptations.

\subsubsection{Invariant Representation Learning}

The module optimizes a generator \(\Phi^*\) to satisfy the invariance condition:
\begin{equation}
P^e(Y | \Phi(\mathcal{M})) = P^{e'}(Y | \Phi(\mathcal{M})),
\label{eq:invariance_condition_theory}
\end{equation}
where \(Y\) represents user-item interactions. The optimization objective is:
\begin{equation}
\Phi^* = \arg \min_{\Phi \in \mathcal{I}_{\mathcal{E}}} E_e(R^e) + \alpha \cdot \text{Var}_e(\nabla_\theta R^e),
\label{eq:optimal_generator}
\end{equation}
where \(R^e = \mathbb{E}_{(u,i) \sim \mathcal{E}^e} [ \ell(\mathbf{h}_u, \mathbf{h}_i, y_{u,i}) ]\) is the risk under environment \(e\), and \(\ell\) is the loss function (\eg BPR loss). The gradient variance term is:
\begin{equation}
\text{Var}_e(\nabla_\theta R^e) = \mathbb{E}_e \left[ \|\nabla_\theta R^e - \mathbb{E}_e[\nabla_\theta R^e]\|_2^2 \right],
\label{eq:gradient_variance}
\end{equation}
penalizing environment-specific fluctuations in the gradients.

The invariant representations are generated via:
\begin{equation}
\mathbf{Z}_I = \text{GNN}^I_{\theta_I}(\mathcal{M}_I, \mathbf{X}), \quad \mathbf{h}_r = \text{READOUT}^I(\mathbf{Z}_I),
\label{eq:invariant_embedding}
\end{equation}
where \(\mathbf{Z}_I\) are node- and graph-level embeddings. The invariance error is bounded by:
\begin{equation}
\mathbb{E}_{e,e'} \left[ |P^e(Y | \mathbf{h}_r) - P^{e'}(Y | \mathbf{h}_r)| \right] \leq \epsilon_{\text{inv}},
\label{eq:invariance_error}
\end{equation}
where \(\epsilon_{\text{inv}}\) decreases as the objective in Equation \eqref{eq:optimal_generator} is minimized.

\section{Optimization Algorithm}
In this section, we outline the algorithmic workflow of our proposed InvGCLLM model, as detailed in Algorithm 1. The process begins with the Causal Score Generation module, which identifies latent environments and trains a generator to produce a causal score matrix $\mathbf{S}$. This matrix quantifies the stability of each user-item interaction. Next, the LLM-Calibrated Causal Graph Editing module leverages these scores to create two distinct graph views: a purified Causal View $\mathcal{G}_c$ and a Spurious View $\mathcal{G}_s$. Finally, these views are used in the Causal-Informed Contrastive Learning stage, where the model is trained with a joint objective that combines our novel contrastive loss $\mathcal{L}_{\text{CICL}}$ and a standard BPR recommendation loss. This end-to-end process yields purified user and item embeddings $\{\mathbf{h}_v\}$ that are robust to distribution shifts.

\subsection{Overview of the InvGCLLM Algorithm}

The optimization algorithm for InvGCLLM, detailed in Algorithm 1, orchestrates the training process. The algorithm takes as input the user-item interaction graph $\mathcal{G}$ and the model components, and its final output is the set of purified user and item embeddings, $\{\mathbf{h}_u\}$ and $\{\mathbf{h}_i\}$.

The process begins by identifying latent environments to guide the learning of the invariant information generator $\Phi(\cdot)$ (Lines 3-5). A dedicated loop then optimizes this generator for $j$ epochs by minimizing the invariant learning objective from Eq. \eqref{eq:objective_function}, resulting in a meaningful causal score matrix $\mathbf{S}$ (Lines 6-9).

Next, the LLM-Calibrated Causal Graph Editing module uses the top-K and bottom-K scores from $\mathbf{S}$ to prompt an LLM. The LLM's adjudications are used to construct the Causal View $\mathcal{G}_c$ and the Spurious View $\mathcal{G}_s$ (Lines 10-11).

Finally, the main training loop runs for $k$ epochs (Lines 12-18). In each iteration, the GNN encoder generates three sets of embeddings: the main representations $\mathbf{h}_v$ (from $\mathcal{G}$), causal representations $\mathbf{h}_{v,c}$ (from $\mathcal{G}_c$), and spurious representations $\mathbf{h}_{v,s}$ (from $\mathcal{G}_s$) (Line 13). These are used to compute the Causal-Informed Contrastive Loss $\mathcal{L}_{\text{CICL}}$ via Eq. \eqref{eq:cicl_loss_final} and the BPR loss $\mathcal{L}_{\text{BPR}}$ via Eq. \eqref{eq:bpr_loss}. The total loss from Eq. \eqref{eq:total_loss} is then used to update the GNN parameters (Lines 14-16). The final, purified embeddings are returned after training is complete (Line 19).

\begin{algorithm}[ht]
\caption{Optimization Algorithm for InvGCLLM}
\label{alg:invgcllm}
\begin{algorithmic}[1]
\State \textbf{Input:} User-item interaction graph $\mathcal{G}$; GNN encoder $f_{\text{GNN}}(\cdot; \Theta)$; Invariant information generator $\Phi(\cdot)$; Large Language Model $f_{\text{LLM}}(\cdot)$; Number of generator epochs $j$; Number of main training epochs $k$.
\State \textbf{Output:} Optimized user and item embeddings $\{\mathbf{h}_u\}$, $\{\mathbf{h}_i\}$.

\State \textit{// Stage 1: Causal Score Generation}
\State Decompose graph into $\mathcal{M}_I, \mathcal{M}_V$ using the generator $\Phi$ by Eq. \eqref{eq:mask_decomposition}.
\State Infer environment labels $\mathcal{E}_{\text{infer}}$ via K-means on embeddings from $\mathcal{M}_V$.
\State Update environment labels $\mathcal{E} \leftarrow \mathcal{E}_{\text{infer}}$.
\For{$epoch = 1, \dots, j$}
    \State Optimize the invariant learning objective from Eq. \eqref{eq:objective_function}.
    \State Update the invariant information generator $\Phi$ to refine the causal score matrix $\mathbf{S}$.
\EndFor

\State \textit{// Stage 2: LLM-Calibrated Causal Graph Editing}
\State Construct Causal View $\mathcal{G}_c$ and Spurious View $\mathcal{G}_s$ based on LLM adjudications of top/bottom-K edges from $\mathbf{S}$.

\State \textit{// Stage 3: Joint Optimization with Causal-Informed Contrastive Learning}
\For{$epoch = 1, \dots, k$}
    \State Generate representations: $\mathbf{h}_v \leftarrow f_{\text{GNN}}(\mathcal{G}, \mathbf{X})$, $\mathbf{h}_{v,c} \leftarrow f_{\text{GNN}}(\mathcal{G}_c, \mathbf{X})$, $\mathbf{h}_{v,s} \leftarrow f_{\text{GNN}}(\mathcal{G}_s, \mathbf{X})$.
    \State Compute contrastive loss $\mathcal{L}_{\text{CICL}}$ by Eq. \eqref{eq:cicl_loss_final}.
    \State Compute recommendation loss $\mathcal{L}_{\text{BPR}}$ by Eq. \eqref{eq:bpr_loss} using purified embeddings $\{\mathbf{h}_v\}$.
    \State Compute total loss $\mathcal{L}_{\text{total}} = \mathcal{L}_{\text{CICL}} + \lambda \mathcal{L}_{\text{BPR}}$ by Eq. \eqref{eq:total_loss}.
    \State Update GNN parameters $\Theta$ by backpropagating $\mathcal{L}_{\text{total}}$.
\EndFor

\State \textbf{return} Final purified user embeddings $\{\mathbf{h}_u\}$ and item embeddings $\{\mathbf{h}_i\}$.
\end{algorithmic}
\end{algorithm}

\subsection{Workflow and Prompts for LLM-Calibrated Causal Graph Editing (LLMC2GE)}

\begin{figure}[h]
\centering
\includegraphics[width=\linewidth]{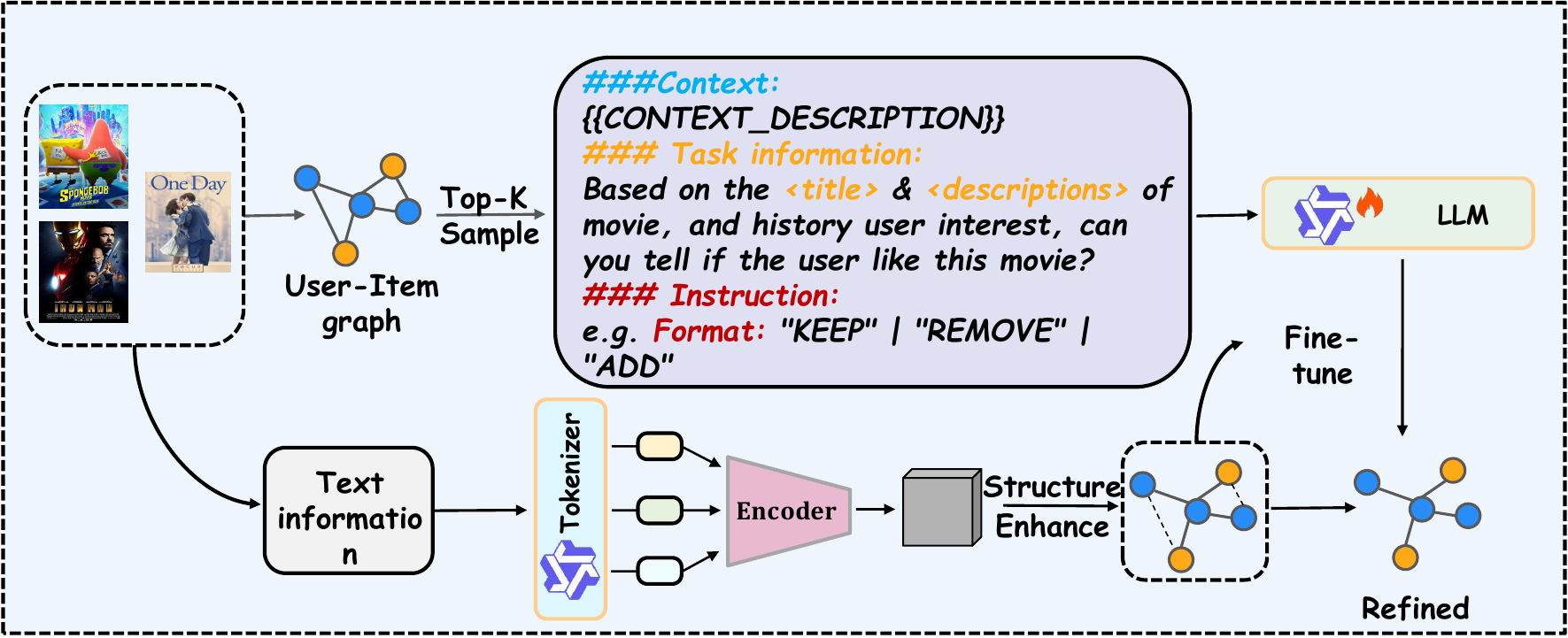}
\caption{The LLMC2GE workflow: The process starts with a prompt that leverages user history and MovieLens-1M data to predict interest, followed by top-$k$ and bottom-$k$ selection via an Edge Predictor, LoRA fine-tuning of the LLM, and graph structure refinement with added edges.}
\label{fig:LLMs}
\end{figure}

In this section, we detail the workflow of the LLM-Calibrated Causal Graph Editing (LLMC2GE) module. This module's purpose is to leverage the reasoning capabilities of a Large Language Model (LLM) to refine the graph structure by verifying the data-driven signals from the causal score matrix $\mathbf{S}$. This process creates high-purity causal and spurious graph views, which are essential for the subsequent contrastive learning stage. The overall workflow is illustrated in Figure~\ref{fig:LLMs}.

The LLMC2GE process does not perform open-ended item recommendation. Instead, it operates as a sophisticated adjudication mechanism. As outlined in Algorithm~\ref{alg:invgcllm} (Line 10), the first step is to manage the computational cost by selecting a focused subset of edges for the LLM to evaluate. We use the causal score matrix $\mathbf{S}$ to identify two candidate sets: a set of potentially causal links, consisting of the top-K highest-scoring edges, and a set of potentially spurious links, consisting of the bottom-K lowest-scoring edges. This strategy ensures that the LLM's powerful, but expensive, reasoning is applied only to the most uncertain or impactful user-item interactions.

\begin{figure}[h]
\centering
\includegraphics[width=\linewidth]{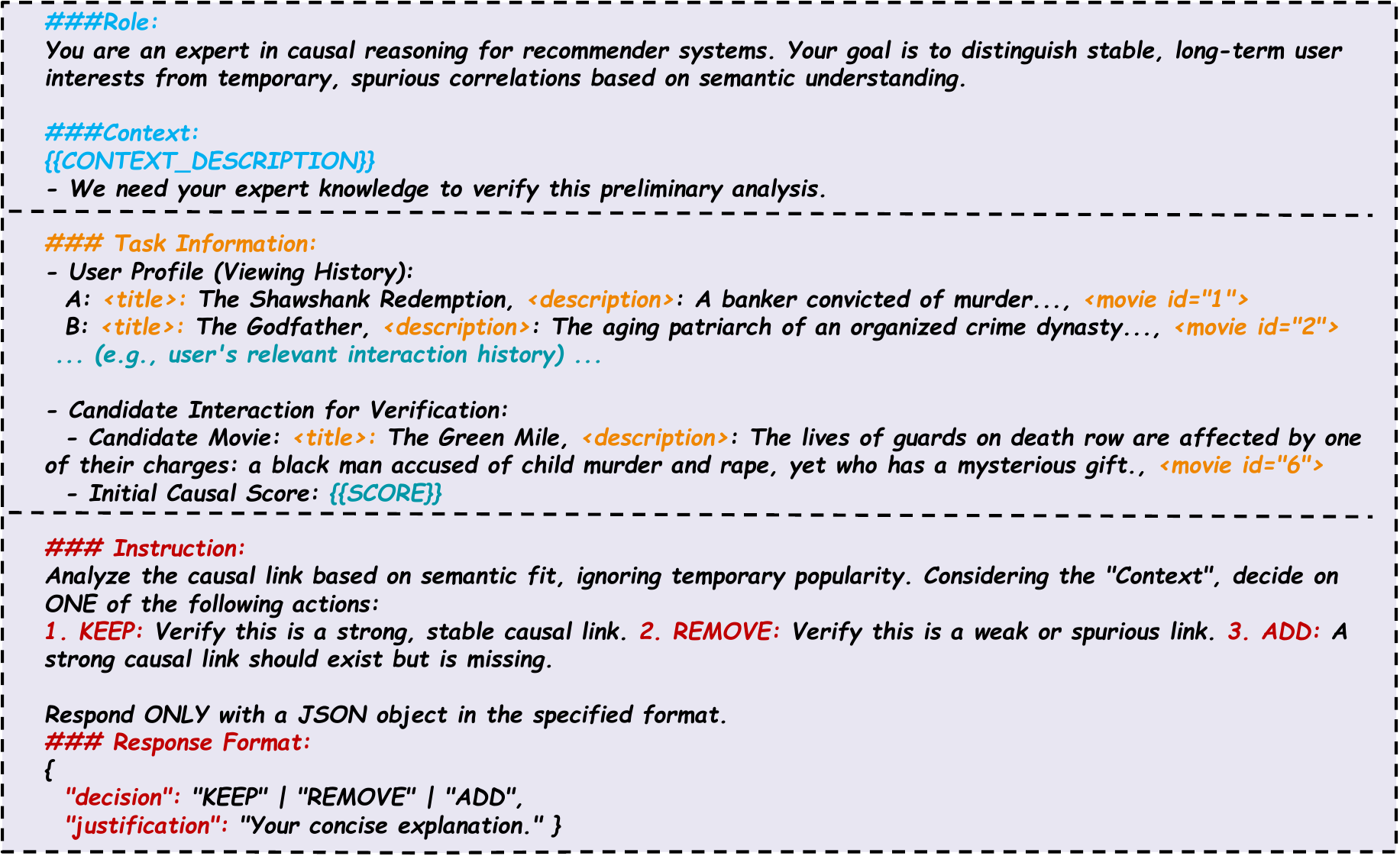}
\caption{The specific prompt used in LLMC2GE, providing the LLM with user viewing history and candidate item descriptions from the MovieLens-1M dataset to predict user interest.}
\label{fig:prompts}
\end{figure}

For each candidate edge, we prompt the LLM to perform a semantic evaluation and make a judgment. The design of the prompt, shown in Figure~\ref{fig:prompts}, is critical. It provides the LLM with rich context, including the user's historical preferences (\eg titles and descriptions of previously liked movies) and the attributes of the candidate item. Crucially, the prompt also includes the data-driven causal score from $\mathbf{S}$ as an additional signal. The LLM's task is to synthesize this information and output a discrete decision, such as confirming an edge as either `CAUSAL` or `SPURIOUS`.

To enhance the LLM's performance on this specific adjudication task, we employ Low-Rank Adaptation (LoRA) for efficient fine-tuning, as depicted in Figure~\ref{fig:LLMs}. The LLM is fine-tuned on a set of representative top-K and bottom-K candidates, which adapts its general world knowledge to the specific nuances of predicting causal relationships in the recommendation domain.

Finally, based on the LLM's decisions, we construct the two refined graph views as defined in the previous section. The Causal View ($\mathcal{G}_c$) is formed by taking the original graph and removing the edges that the LLM has confirmed as spurious. Concurrently, the Spurious View ($\mathcal{G}_s$) is composed exclusively of those edges that the LLM has identified and confirmed as spurious. These two purified graphs, $\mathcal{G}_c$ and $\mathcal{G}_s$, are then passed to the Causal-Informed Contrastive Learning module. This workflow ensures that the graph structures are grounded in both quantitative, data-driven analysis and deep semantic reasoning, strengthening the entire InvGCLLM framework's ability to handle OOD scenarios.

\begin{table*}[t]
\centering
\renewcommand{\arraystretch}{2}
\fontfamily{ptm}\selectfont
\resizebox{\textwidth}{!}{
\begin{tabular}{l|ccc|ccc|ccc|ccc}
\hline
 & \multicolumn{3}{c|}{Douban} & \multicolumn{3}{c|}{Amazon-Book} & \multicolumn{3}{c|}{MovieLens-1M} & \multicolumn{3}{c}{Yahoo} \\
\cline{2-13}
 & NDCG@K & Pre@K & Recall@K & NDCG@K & Pre@K & Recall@K & NDCG@K & Pre@K & Recall@K & NDCG@K & Pre@K & Recall@K \\
\hline
LightGCN & 0.0781 & 0.0534 & 0.0745 & 0.0235 & 0.0142 & 0.0291 & 0.2057 & 0.1768 & 0.1349 & 0.0749 & 0.0156 & 0.1529 \\
LightGCL & 0.0763 & 0.0487 & 0.0739 & 0.0259 & 0.0162 & 0.0321 & 0.2079 & 0.1785 & 0.1367 & 0.0752 & 0.0137 & 0.1494 \\
SGL & 0.0785 & 0.0471 & 0.0744 & 0.0258 & 0.0156 & 0.0312 & 0.2186 & 0.1854 & 0.1425 & 0.0763 & 0.0134 & 0.1571 \\
InvCF & 0.0779 & 0.0478 & 0.0734 & 0.0237 & 0.0136 & 0.0289 & 0.2161 & 0.1792 & 0.1373 & 0.0757 & 0.0145 & 0.1563 \\
BOD & 0.0834 & 0.0436 & 0.0697 & 0.0267 & 0.0172 & 0.0327 & 0.2149 & 0.1742 & 0.1361 & 0.0824 & 0.0129 & 0.1492 \\
APDA & 0.0796 & 0.0574 & 0.0671 & 0.0246 & 0.0149 & 0.0304 & 0.2382 & 0.1987 & \underline{0.1536} & 0.0771 & 0.0148 & 0.1599 \\
GraphDA & \underline{0.0937} & 0.0619 & \underline{0.0835} & 0.0218 & 0.0131 & 0.0287 & 0.2235 & 0.1884 & 0.1479 & 0.0766 & 0.0149 & 0.1542 \\
DR-GNN & 0.0913 & \underline{0.0732} & 0.0823 & \underline{0.0339} & \underline{0.0198} & \underline{0.0419} & \underline{0.2362} & \underline{0.1947} & 0.1493 & \underline{0.0789} & \underline{0.0141} & \underline{0.1644} \\
\rowcolor{gray!25} \textbf{InvGCLLM} & \textbf{0.1269*} & \textbf{0.0873*} & \textbf{0.0942*} & \textbf{0.0432*} & \textbf{0.0236*} & \textbf{0.0445*} & \textbf{0.2547*} & \textbf{0.2139*} & \textbf{0.1578*} & \textbf{0.0948*} & \textbf{0.0162*} & \textbf{0.1754*} \\
\hline
\end{tabular}
}

\caption{Overall performance on four different datasets. Bold numbers indicate the highest scores, while underlined numbers represent the second-best results among all. Adding ``\textbf{{\Large *}}'' denotes the models have statistically significant improvements (i.e., two-sided t-test with $p<0.05$).}

\label{tab:performance}
\end{table*}
\section{Experiments}

In this section, we perform extensive experiments to assess the effectiveness of the proposed framework. To provide a clear analysis, the experiments are designed to address the following research questions (\textbf{RQs}):

\begin{itemize}[leftmargin=*]
\item {\textbf{RQ1}}: How does InvGCLLM perform compared to current state-of-the-art (SOTA) benchmark models?
\item {\textbf{RQ2}}: How does LLM-based contrastive learning compare to traditional contrastive learning?
\item {\textbf{RQ3}}: How do hyperparameters influence the performance of our model?
\item {\textbf{RQ4}}: What are the contributions of the individual components of InvGCLLM to overall performance?
\item {\textbf{RQ5}}: How robust is each of the models we compare to distribution shifts?
\end{itemize}

We begin by introducing the experimental setup, followed by an exploration of the answers to these four questions.

\subsection{Experimental Setting}
In this section, we describe the experimental settings employed to evaluate the performance of various models under different types of distribution shifts. 

\subsubsection{Datasets}
Our experiments encompass three types of shifts: \textit{popularity shift}, \textit{temporal shift}, and \textit{exposure shift}, followed by~\cite{wang2024distributionally}. For the popularity shift, we utilized the Douban\footnote{\url{https://www.kaggle.com/datasets/}} and Amazon-Book\footnote{\url{https://jmcauley.ucsd.edu/data/amazon/}} datasets, re-partitioning the training and test sets based on item popularity, such that the test set was designed to have a near-uniform distribution of item popularity. For the temporal shift, we used the MovieLens-1M dataset\footnote{\url{https://grouplens.org/datasets/movielens/}}. In this setup, the most recent 20\% of interaction data was used for testing, while the earliest 60\% was used for training. In the exposure shift setting, we constructed the training set based on ratings collected by the actual system, while the test set consisted of randomly selected item ratings, resulting in a biased training set and an unbiased test set. This setup was experimented on the Yahoo Music dataset~\cite{schnabel2016recommendations}. 

\subsubsection{Metrics}

For evaluation, we employed three widely used metrics in RS: NDCG@K, Precision@K, and Recall@K, with K being set by default as 10. NDCG@K measures the ranking quality of the recommended items. Precision@K Indicates the proportion of relevant items among the recommended ones. Recall@K reflects the model's ability to retrieve all relevant items. Higher values mean higher performance.

\subsubsection{Baselines}
The baseline methods we compared are categorized into three main types: robustness methods for recommender systems, graph contrastive learning methods, and reconstruction-based methods. Notably, the methods we studied are as follows: InvCF and BOD belong to robustness methods for recommender systems. LightGCL and SGL are graph contrastive learning methods. Both APDA and GraphDA are reconstruction-based methods. The performance of these methods on the datasets is detailed in Table~\ref{tab:performance}:
\begin{itemize}[leftmargin=*]

\item \textbf{LightGCN \cite{he2020lightgcn}:} A simplified Graph Convolutional Network (GCN) that removes feature transformation and non-linear activation layers. Its streamlined design focuses on linear neighborhood aggregation to effectively capture high-order connectivity in the user-item interaction graph, achieving both accuracy and efficiency in recommendations.

\item \textbf{LightGCL \cite{cai2023lightgcl}:} LightGCL is a graph contrastive learning method. It leverages Singular Value Decomposition (SVD) to enhance the intrinsic structure of the graph. By effectively utilizing the preserved structure, it demonstrates strong performance in recommendation.

\item \textbf{SGL \cite{wu2021self}:} SGL is a foundational graph contrastive learning method. It constructs positive samples to enhance the robustness and performance of recommendation models through contrastive learning.

\item \textbf{InvCF \cite{zhang2023invariant}:} InvCF is a robustness method aimed at addressing distribution shift issues in recommender systems. It tackles popularity bias through invariant learning, ensuring the model remains robust across different data distributions.

\item \textbf{BOD \cite{wang2023efficient}:} BOD is another robustness method. It employs bilevel optimization to achieve data denoising, enhancing the model's performance in handling noisy data.

\item \textbf{APDA \cite{zhou2023adaptive}:} APDA enhances generalization by continuously reducing the edge weights of unpopular items in user-item graphs while increasing the edge weights of popular items. This approach has proven particularly effective in addressing out-of-distribution issues in recommender systems.

\item \textbf{GraphDA~\cite{fan2023graph}:} GraphDA employs a pre-trained encoder to reconstruct the adjacency matrix, thereby amplifying signals and reducing noise within the graph.

\item \textbf{DR-GNN \cite{wang2024distributionally}:} DR-GNN is a SOTA robustness model designed to handle distribution shift issues using a Distributionally Robust Optimization (DRO) framework. It ensures robust performance across shifts that could potentially impact recommendation accuracy by optimizing the model over various distributions.
\end{itemize}

\subsubsection{Implementation Details.}
The models were trained for a total of 100 epochs with a batch size of 256. The learning rate was set to 0.0001, which was determined through preliminary experiments to ensure optimal convergence. This model uses the conventional LightGCN as the backbone and BPR loss. The LLM is fine-tuned using Low-Rank Adaptation (LoRA) to suit the graph recommendation's data better. In this paper, we use \textit{Qwen3-8B} for our experiments. 

\subsection{Overall Performance (RQ1)}

To address \textbf{RQ1}, we compared InvGCLLM with representative benchmark models. The comparison results are summarized in Table~\ref{tab:performance}, where models such as LightGCN, SGL, and GAT represent various SOTA techniques in graph-based recommender systems, while methods like InvCF, BOD, and DR-GNN focus on addressing distribution shift issues.

From Table~\ref{tab:performance}, several key observations can be made:
(i) Firstly, our method, InvGCLLM, consistently outperforms all other methods on the Douban, Amazon-Book, and Yahoo datasets, achieving the highest NDCG, Precision, and Recall scores. The superior performance of InvGCLLM is attributed to its innovative framework that combines the Invariant Learning Module with graph contrastive learning, enabling the model to better adapt to changes in data distribution.
(ii) Methods specifically designed to address distribution shift, such as BOD, InvCF, and DR-GNN, demonstrate significant performance improvements on the Douban and Amazon-Book datasets compared to standard graph methods like LightGCN and SGL. This underscores the importance of considering distribution robustness in recommender systems.
(iii) LightGCL, which incorporates contrastive learning, shows a slight improvement over LightGCN, indicating that contrastive learning can enhance the model's ability to capture subtle user-item interactions.

In summary, the outstanding performance of InvGCLLM across multiple datasets, particularly its superiority on the Douban, Amazon-Book, and Yahoo datasets, validates the effectiveness of our proposed method. InvGCLLM not only successfully addresses the challenges posed by distribution shifts but also surpasses existing SOTA models, establishing itself as a highly robust and effective solution in graph-based recommender systems.

\subsection{Ablation Study (RQ2, RQ4)}

\begin{figure}
	\centering
	\includegraphics[width=\linewidth]{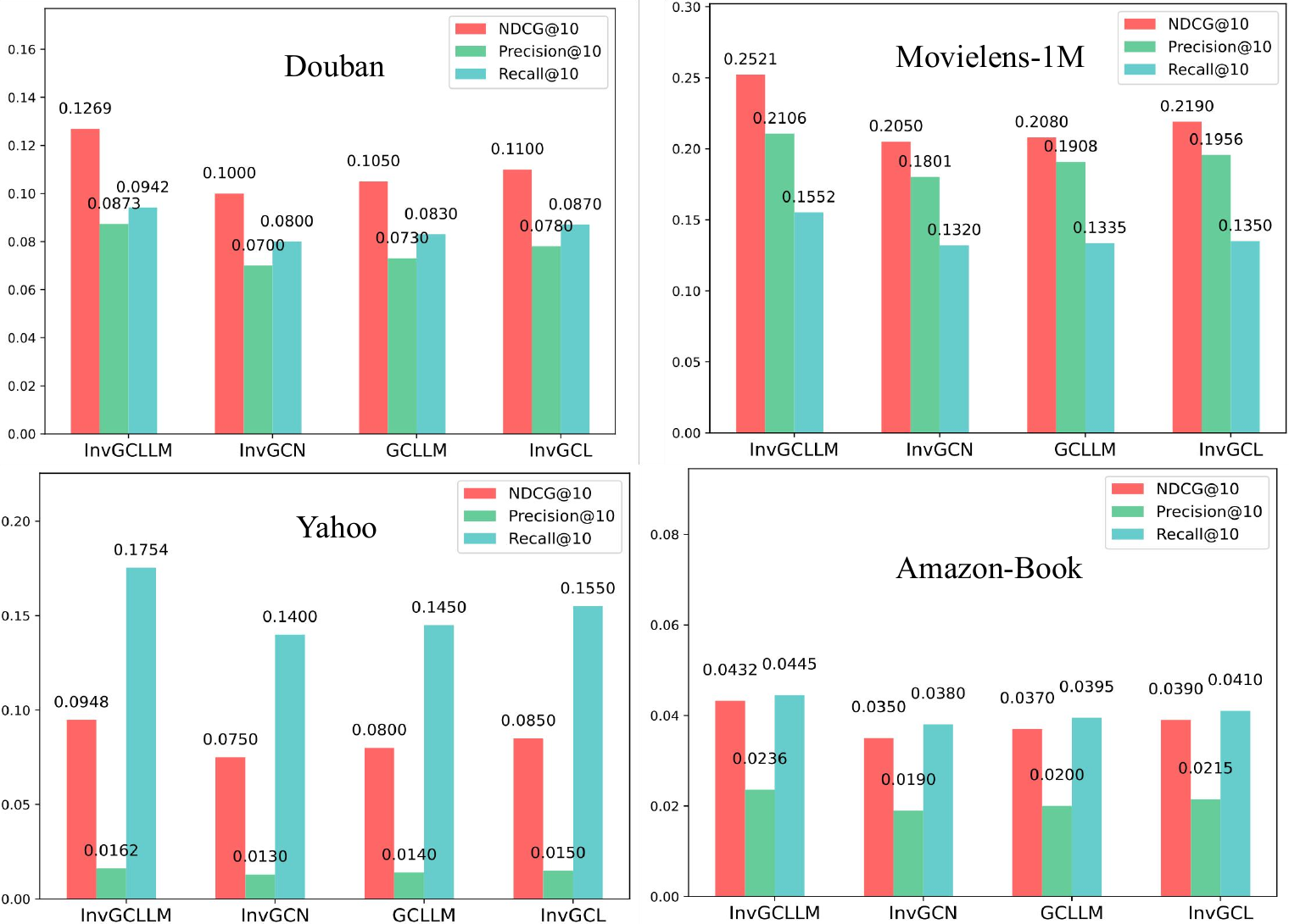}
	\caption{Ablation study results on four datasets (Douban, Movielens-1M, Yahoo, Amazon-Book), comparing InvGCLLM, InvGCN, GCLLM, and InvGCL in terms of NDCG@10, Precision@10, and Recall@10. InvGCLLM consistently achieves the best performance.}\label{fig:abla}
\end{figure}

To answer \textbf{RQ2} and \textbf{RQ4}, we performed an ablation study comparing in all four datasets. To be specific, our ablation study is structured into three key parts. First, we retain the original LightGCN framework, remove the LLM-based contrastive learning, and only keep the invariant module, resulting in a model termed InvGCN. Second, we remove the invariant module and retain only the LLM-based contrastive learning, with LightGCN as the backbone, naming this model GCLLM. Third, we replace the contrastive learning in InvGCLLM with traditional contrastive learning, which requires manual selection of positive and negative samples, and refer to this model as InvGCL. The experimental results are illustrated in the Figure~\ref{fig:abla}.
\begin{itemize}[leftmargin=*]
    \item \textbf{InvGCLLM vs. InvGCN:} This comparison evaluates the impact of leveraging large language models (LLMs) in contrastive learning settings. The results show that InvGCLLM significantly outperforms InvGCN across all metrics (NDCG@K, Precision@K, and Recall@K). This demonstrates that integrating LLM-based contrastive learning into the model enhances its ability to capture complex user-item interactions, leading to more accurate recommendations. The inclusion of LLMs provides richer contextual understanding, which is particularly beneficial in recommendation scenarios.
    
    \item \textbf{InvGCLLM vs. GCLLM:} This analysis examines the effect of incorporating invariance principles into contrastive learning. The results indicate that InvGCLLM achieves higher scores compared to GCLLM, suggesting that the introduction of invariance principles enhances the model’s robustness and its ability to generalize across different data distributions. The improved performance highlights the importance of combining invariance principles with contrastive learning, leading to a more stable and effective recommendation system.
    
    \item \textbf{InvGCLLM vs. InvGCL:} This comparison assesses the benefit of our proposed contrastive learning method over traditional methods. InvGCLLM outperforms InvGCL in all metrics, emphasizing that our approach, which avoids manual selection of positive and negative samples, is more efficient and less prone to errors. This result underscores the advantage of using a more automated and integrated approach to contrastive learning, which better captures the underlying patterns in the data without the need for extensive manual intervention.
\end{itemize}

In summary, InvGCLLM consistently delivers superior performance across all comparisons, validating the effectiveness of our approach in enhancing recommendation accuracy and robustness. The results underscore the importance of both LLM-based contrastive learning and the integration of invariance principles, establishing InvGCLLM as the most effective model in this ablation study.

\begin{figure}
	\centering
	%	\hspace*{-0.6cm}
	{\subfigure{\includegraphics[width=0.49\linewidth]{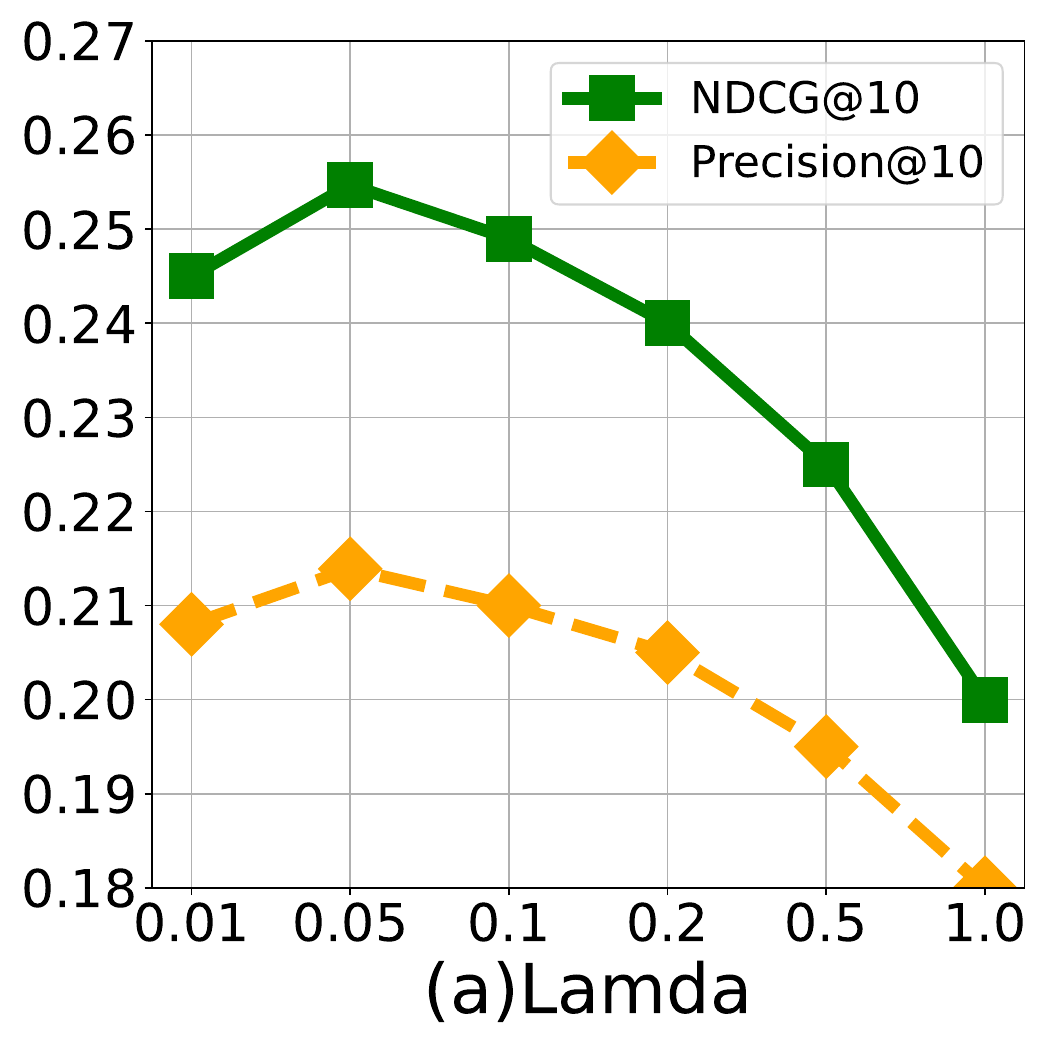}}}
	{\subfigure{\includegraphics[width=0.49\linewidth]{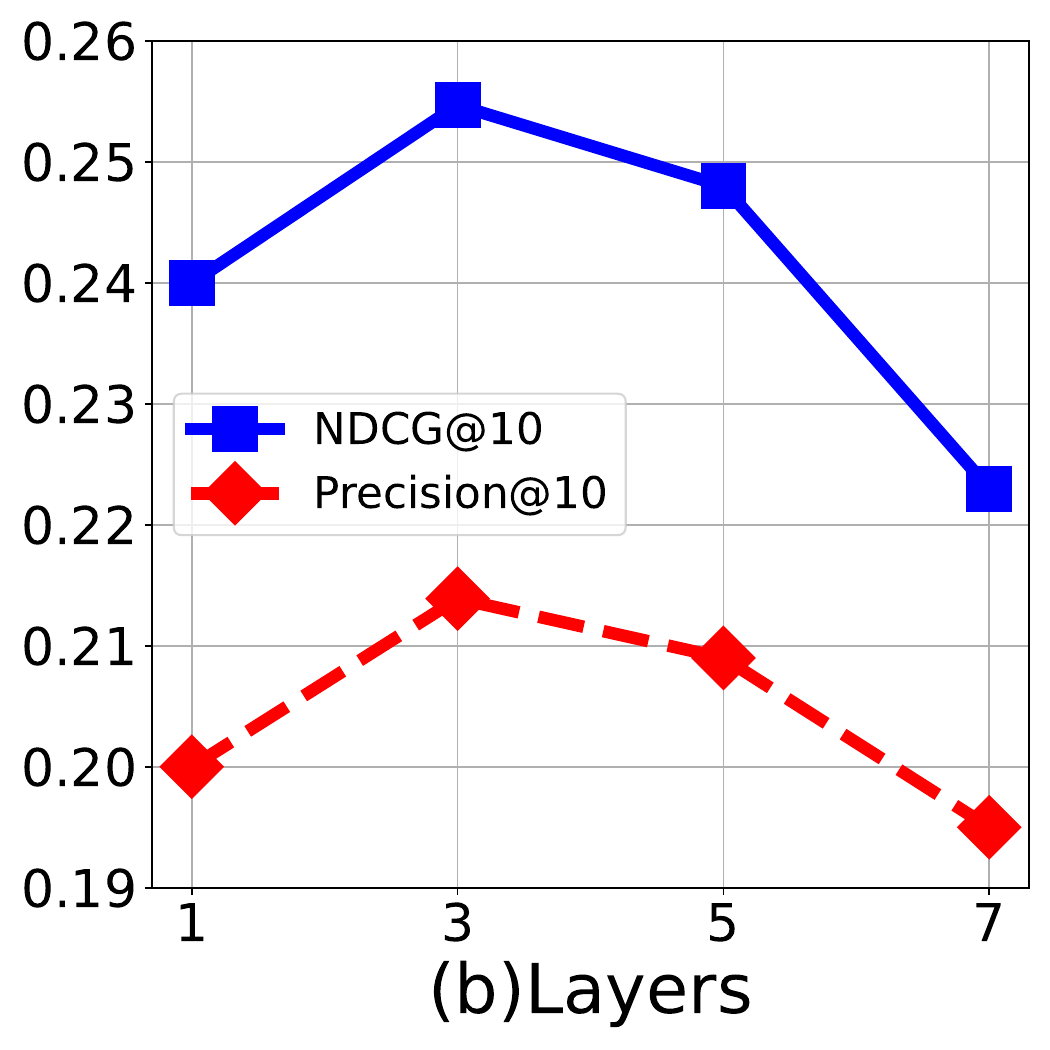}}}
% 	\vspace*{-2mm}
	\caption{Parameter sensitivity analysis on Movielens-1M. (a) Model performance with different $\lambda$ values for balancing recommendation and contrastive losses. (b) Model performance with varying numbers of GNN layers.}\label{fig:para}
% 	\vspace{-2.9mm}
\end{figure}

\subsection{Parameter Analysis (RQ3)}

To answer \textbf{RQ3}, we further investigated the sensitivity of our model to various hyperparameters in Movielens-1M:
\begin{itemize}[leftmargin=*]
    \item \textbf{Lambda ($\lambda$):} $\lambda$ is a hyperparameter that balances the loss of the recommendation system and the loss of graph contrastive learning. The range of $\lambda$ values tested included [0.01, 0.05, 0.1, 0.2, 0.5]. As shown in Figure~\ref{fig:para}(a), the model achieved optimal performance on Movielens-1M when $\lambda = 0.05$ on Movielens-1M. When $\lambda$ is small (\eg from 0.01 to 0.05), the model tends to underperform, as it focuses too heavily on structural information and does not effectively utilize interaction data, resulting in suboptimal recommendation accuracy. However, when $\lambda$ exceeds 0.05, the performance declines sharply. This is likely because the model becomes overly focused on interaction data, neglecting the graph's structural information, leading to overfitting and poor generalization.

    \item \textbf{Layers:} The range of layers tested included [1, 3, 5, 7]. As shown in Figure~\ref{fig:para}(b), the best performance was observed with three layers. A single layer was often too simplistic to capture the intricate relationships in the data, leading to suboptimal results. However, increasing the number of layers beyond three tended to introduce excessive complexity, which not only increased the risk of overfitting but also led to higher computational costs without significant performance gains. Therefore, a carefully chosen number of layers is crucial for achieving both efficient and effective model performance.
\end{itemize}

From these observations, we can conclude that:
(i) The results indicate that parameter $\lambda$ plays a critical role in balancing contrastive learning and interaction-based learning. Therefore, selecting an appropriate $\lambda$ is crucial to maintaining a balance between structural learning and interaction-based learning, ensuring strong generalization across different datasets.
(ii) Limiting the number of layers to two or three provides sufficient depth for capturing complex patterns without introducing unnecessary computational overhead or the risk of overfitting.

\subsection{Visualization Analysis (RQ5)}\label{sec:visualization}
\begin{figure*}
	\centering
	\includegraphics[width=\textwidth]{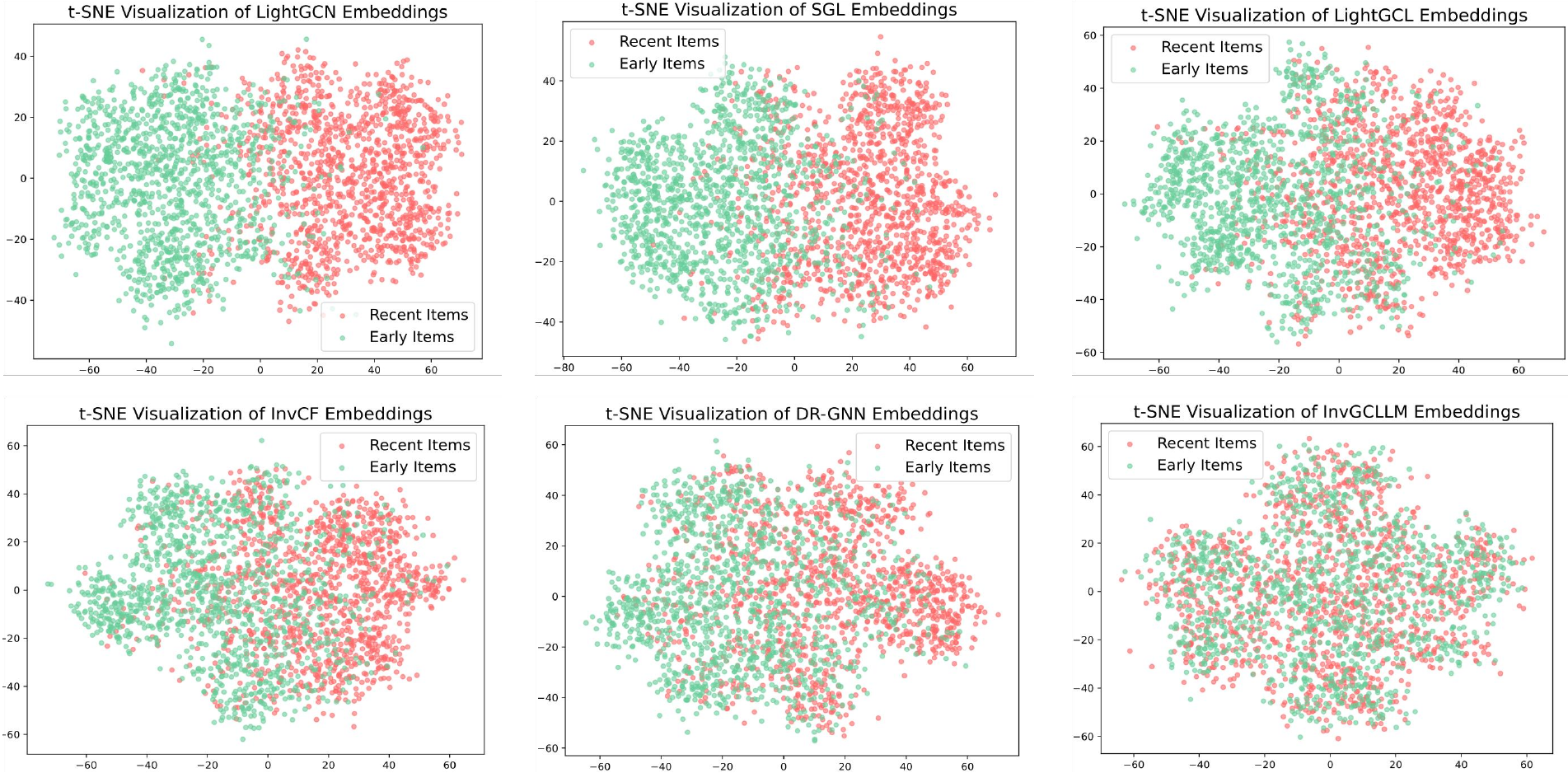}
	\caption{t-SNE visualization of item representations on Movielens-1M under temporal distribution shift. Compared to baseline models, InvGCLLM yields the most uniform embedding distribution across recent and early items, demonstrating superior robustness and generalization to temporal shifts.}\label{fig:tsne}
\end{figure*}

To answer \textbf{RQ5}, we conduct a visualization analysis to evaluate the robustness of different models under distribution shift. Specifically, we use the Movielens-1M dataset, which is subject to temporal shift, and select the most recent 20\% and the earliest 20\% of the interaction data. As shown in Figure~\ref{fig:tsne}, we visualize the item representations learned by six models: LightGCN, SGL, LightGCL, InvCF, DR-GNN, and our proposed InvGCLLM.

(i) First, we observe that the embeddings learned by LightGCN exhibit the most obvious separation between recent and early items, with two distinct clusters forming in the representation space. This clear boundary indicates that LightGCN suffers from poor generalization under temporal distribution shift.

(ii) For InvCF and DR-GNN, the embeddings of recent and early items are more mixed within the same space, resulting in less separation between the two segments. This phenomenon demonstrates that robustness-based methods can mitigate the impact of distribution shift, thereby improving OOD generalization.

(iii) The embeddings of LightGCL are also more evenly distributed compared to LightGCN and closely resemble those of InvCF and DR-GNN, suggesting that GCL-based approaches possess a certain degree of out-of-distribution robustness.

(iv) Notably, the embeddings produced by our InvGCLLM are the most uniformly and evenly distributed across both recent and early items, with no obvious separation. This more uniform distribution within the same space indicates that InvGCLLM achieves the best generalization ability and is most effective in handling temporal distribution shifts among all compared models.

\section{Related Work}

\subsection{Contrastive Learning in Graph Recommendation}
%图推荐系统中的对比学习由于其通过对比正样本和负样本来学习鲁棒特征表示的能力而获得了极大的关注。这种技术尤其适用于图形数据，因为图形数据的复杂结构和互联性带来了独特的挑战。一些研究表明，对比学习可以通过捕获图结构中的关键特征来提高推荐系统中图神经网络(gnn)的性能。尽管取得了这些进步，但这些方法并不是专门为解决OOD (out- distribution)问题而设计的，并且在处理out- distribution数据时往往表现不佳。图数据中的复杂关系和推荐系统对泛化的需求进一步加剧了这一问题。而本文提出的LLM-based的对比学习，通过对图结构的数据增强，可以更好的捕获图数据中的复杂关系，从而提升泛化性能。

Contrastive learning in graph-based recommender systems~\cite{jin2021automated,jin2022empowering,fan2022graph,SHI2025113766} has attracted considerable attention for its effectiveness in learning robust feature representations by contrasting positive and negative samples. This approach is particularly valuable in graph data, where complex structures and high interconnectivity present unique challenges~\cite{ding2022data}. Numerous studies~\cite{yang2024empirical,wu2021self,cai2023lightgcl,LIU2025113217} have shown that contrastive learning can improve the performance of graph neural networks (GNNs) in recommender systems by capturing salient features within graph structures. For instance, SimGCL~\cite{yu2022graph} augments data by adding random noise directly to embedding representations, while LightGCL~\cite{cai2023lightgcl} leverages singular value decomposition (SVD) to guide self-supervised signals. APDA~\cite{zhou2023adaptive} enhances generalization by dynamically reducing the edge weights of frequently selected items and increasing those of less frequently selected items. RGCF~\cite{tian2022learning} promotes data diversity while identifying reliable information transmission interactions. Despite these advancements, existing methods are not explicitly designed to address OOD issues and therefore often struggle to handle OOD data. This limitation stems from their inability to fully capture the complexities of graph data and the generalization requirements of recommender systems in the presence of distribution shifts. 

\subsection{LLM Models in Recommendation}
%在推荐系统中，基于大规模语言模型(LLM)的推荐算法近年来取得了显著的进展。传统的推荐系统方法，如协同过滤和基于内容的推荐，通常依赖于用户和项目的静态特征以及明确的用户反馈。然而，这些方法在处理动态和稀疏数据时存在局限性。近年来，基于大规模语言模型的方法通过捕获用户和项目之间更微妙和复杂的关系，有效地提高了推荐系统的性能。例如，\cite{llmwithgraph}提出GaCLLM通过结合图卷积网络来捕获用户和项目之间的复杂关系。\cite{enhancingrec}提出LLMRG，通过图推理机制提高推荐的准确性。另一项研究\cite{llmrec}提出了LLMRec，通过融合多模态数据提供了更全面的用户画像，从而提高了推荐的有效性。尽管大规模语言模型具有自然的泛化能力，但这些面向Rec的LLM模型并没有专门解决OOD问题。本文针对这一问题提出了一种新的解决方案，通过在推荐系统中引入不变性原则，增强了模型在不同分布下的泛化能力，有效地解决了分布偏差的挑战。
In recent years, recommendation algorithms based on large language models (LLMs) have made significant progress. Traditional recommendation methods~\cite{zhao2022mae4rec,liu2022rating,zheng2022cbr}, typically rely on static user and item features as well as explicit user feedback. However, these approaches often struggle with dynamic and sparse data. LLM-based methods, by capturing more subtle and complex relationships between users and items, have shown great potential in addressing these challenges. For instance, GaCLLM~\cite{du2024large} leverages Graph Convolutional Networks to model complex user-item interactions, while LLMRG~\cite{wang2024llmrg} enhances recommendation accuracy through graph reasoning mechanisms. Additionally, LLMRec~\cite{wei2024llmrec} integrates multimodal data to build more comprehensive user profiles, thereby improving recommendation performance. RPP~\cite{mao2025reinforced} focuses on individual user differences by decomposing prompts into four distinct patterns and combining them through a multi-agent clipping strategy. These personalized prompts iteratively interact with the LLM, allowing for adaptive refinement and improved recommendation performance. G-Refer~\cite{li2025g} integrates graph-based retrieval with large language models to generate explainable recommendations by retrieving and translating collaborative filtering signals into human-understandable explanations. Although the inherent generalization capability of large language models allows these LLM-based recommendation models to excel in recommendation tasks, they do not explicitly address the out-of-distribution (OOD) problem. 

\subsection{OOD problem in recommendation}

Model training typically assumes that the training and test data are drawn from the same distribution; when this assumption is violated, resulting in an out-of-distribution (OOD) issue, model performance can degrade significantly. InvCF~\cite{zhang2023invariant} addresses prevalence bias by introducing an auxiliary classifier that bases recommendations on prevalence. InvPref~\cite{wang2022invariant} and HIRL~\cite{zhang2023hierarchical} mitigate OOD effects by assigning environment variables to each interaction, partitioning the dataset into multiple environments, and employing invariant learning to identify consistent elements across these environments. Causal learning methods, such as COR~\cite{wang2022causal}, utilize causal graph modeling and counterfactual reasoning to address distributional shifts, while BOD~\cite{wang2023efficient} applies a two-layer optimization scheme to assign interaction weights and reduce noise. DT3OR~\cite{yang2025dual} introduces a model adaptation mechanism during the testing phase by designing self-distillation and comparison tasks, enabling the model to learn users’ invariant interest preferences as well as evolving user and item characteristics. CausalDiffRec~\cite{zhao2025graph} employs backdoor adjustment and variational reasoning to infer the true environmental distribution, which is then used as prior knowledge to guide representation learning during the reverse stage of the diffusion process, thereby facilitating the learning of invariant representations. However, these approaches are not specifically tailored for GNN-based recommendation systems and thus do not explicitly address the impact of distribution shifts on the structure of graph models.

\section{Conclusion}
This work presents InvGCLLM, an innovative framework that integrates the Invariant Learning Module with GCL to enhance the robustness and generalization of LLMs in graph-structured data. Extensive experiments on four public datasets indicate InvGCLLM's superior performance in OODRS, surpassing conventional graph learning methods and existing invariant learning approaches.

\section{Acknowledgements}
This work was supported in part by the National Natural Science Foundation of China (62476101, U21A20478, 62222603, 62076102), in part by the National Key R\&D Program of China (2023YFA1011601), in part by the Key-Area Research and Development Program of Guangdong Province under number 2023B0303030001, and in part by the Guangzhou Science and Technology Plan Project (2024A04J3749) and the Fundamental Research Funds for the Central Universities (2024ZYGXZR062), in part by the Young Talent Support Project of Guangzhou Association for Science and Technology (QT-2025-016); the Natural Science Foundation of Guangdong Province (2024A1515010120).

\bibliographystyle{ieeetr}

\bibliography{Reference}

\begin{IEEEbiography} [{\includegraphics[width=1in,height=1.25in,clip,keepaspectratio]{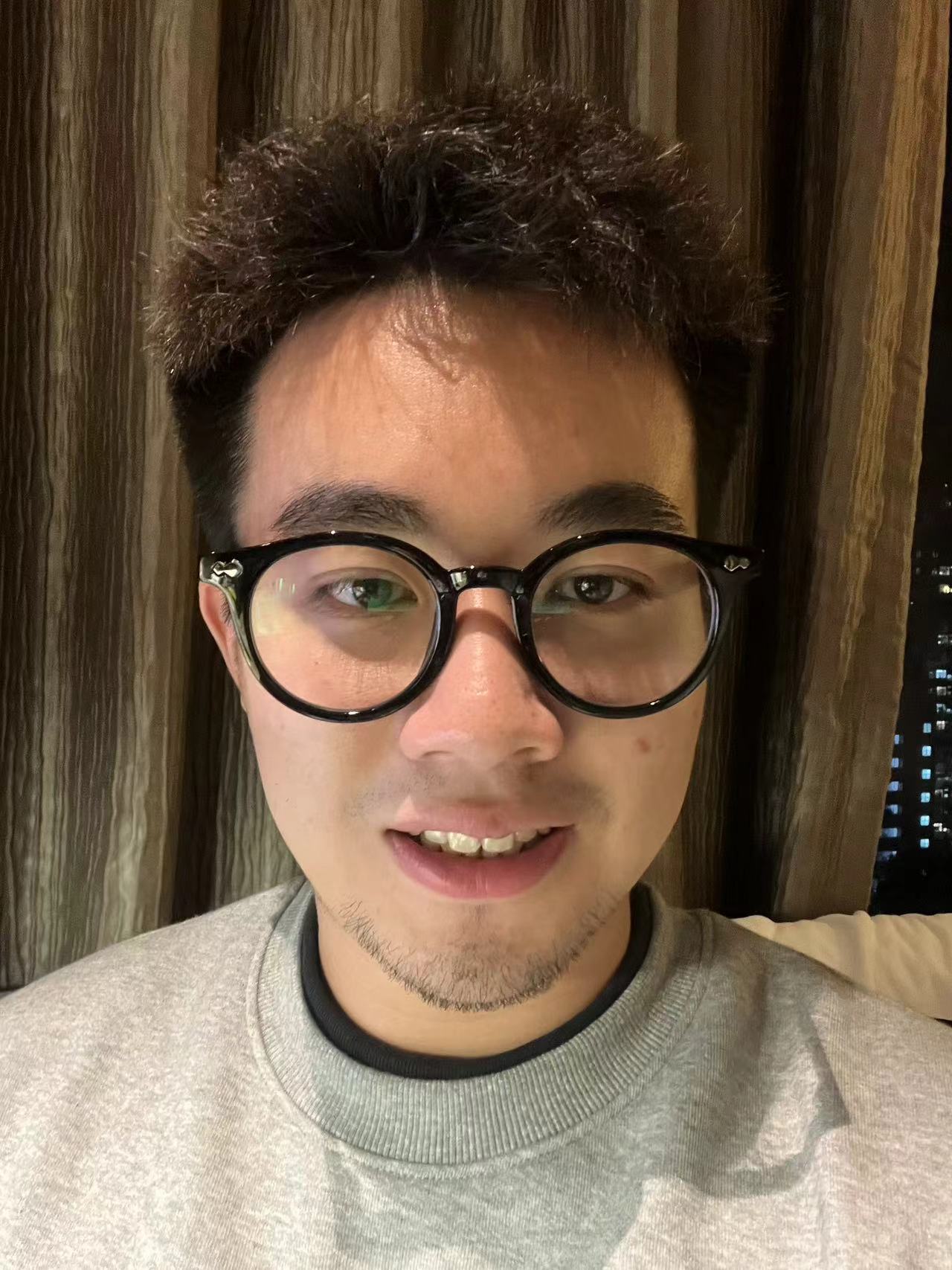}}] {Jiahao Liang} received the M.S. degree from City university of Hong Kong in 2023. He is currently pursuing the Ph.D. degree at the School of Future Technology, South China University of Technology, China. His research interests include machine learning and graph data mining.
\end{IEEEbiography}

\begin{IEEEbiography}[{\includegraphics[width=1in,height=1.25in,clip,keepaspectratio]{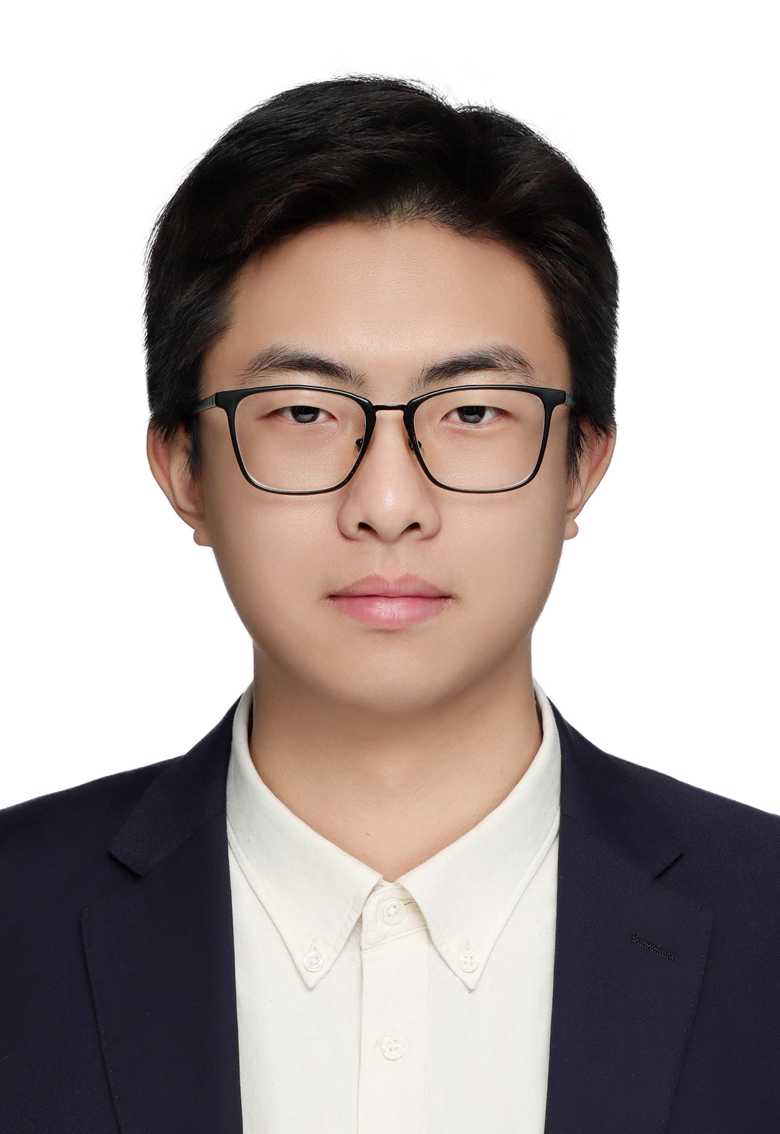}}]{Haoran Yang}
is an Associate Professor of the School of Computer Science and Engineering at Central South University. He obtained his Ph.D. degree at the School of Computer Science of the University of Technology Sydney in 2025, supervised by Prof. Guandong Xu. He received his B.Sc. in Computer Science and Technology and B.Eng. Minor in Financial Engineering from Nanjing University in 2020. His research interests include, but are not limited to, graph data mining and its applications.
\end{IEEEbiography}

\begin{IEEEbiography}[{\includegraphics[width=1in,height=1.25in,clip,keepaspectratio]{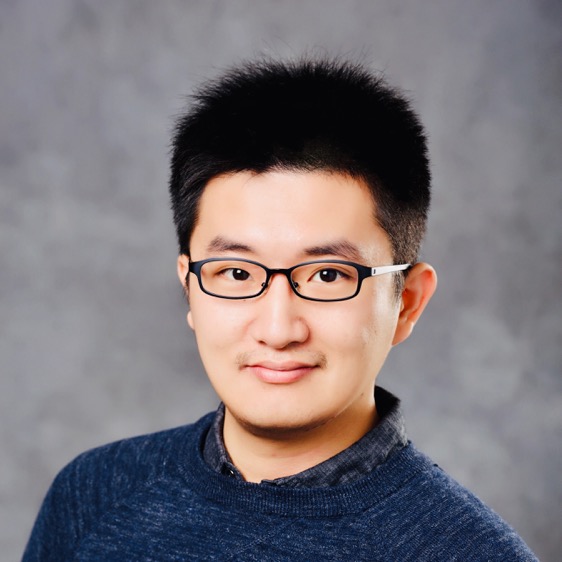}}]{Xiangyu Zhao} is a tenured associate professor (effective from July 2025) of Data Science at City University of Hong Kong (CityU). He worked as an assistant professor at CityU from Sep 2021, and then got an early promotion with tenure. Prior to CityU, he completed his Ph.D. under Prof. Jiliang Tang at MSU, his M.S. under Prof. Enhong Chen at USTC, and his B.Eng. under Prof. Tao Zhou and Prof. Ming Tang at UESTC. He has published more than 100 papers in top conferences and journals. His research has been awarded ICDM'22 and ICDM'21 Best-ranked Papers and Global Top 25 Chinese New Stars in AI (Data Mining). He is a member of the founding academic committee of MLNLP, the largest Chinese AI community with millions of subscribers.
\end{IEEEbiography}

\begin{IEEEbiography} [{\includegraphics[width=1in,height=1.25in,clip,keepaspectratio]{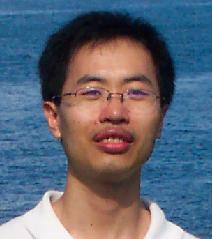}}] {Zhiwen Yu} is a Professor in School of Computer Science and Engineering, South China University of Technology, China.  Dr. Yu serves as an associate editor of IEEE Transactions on Systems, Man, and Cybernetics: Systems. He received the PhD degree from City University of Hong Kong in 2008. Dr. Yu has published more than 150 referred journal papers and international conference papers, including 50 IEEE Transactions papers.
\end{IEEEbiography}

\begin{IEEEbiography} [{\includegraphics[width=1in,height=1.25in,clip,keepaspectratio]{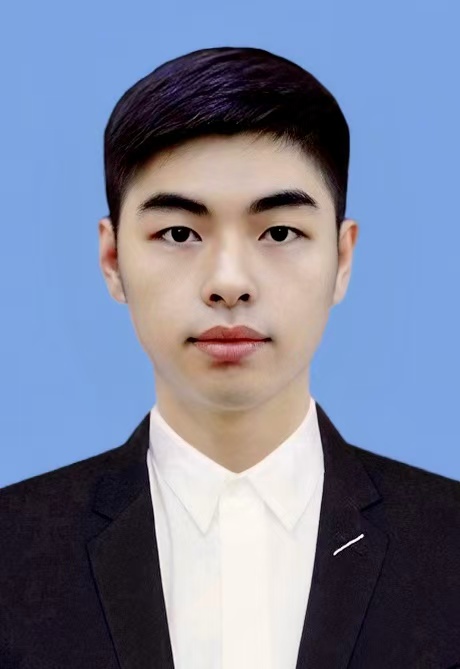}}]{Mianjie Li} received the Ph.D. in Electronics and Information Technology from Macau University of Science and Technology, Macau, China, in 2020. He is currently a Professor with the School of Electronics and Information, Guangdong Polytechnic Normal University, Guangzhou, China. His research interests include image processing, digital multimedia processing and information security.
\end{IEEEbiography}

\begin{IEEEbiography}[{\includegraphics[width=1in,height=1.25in,clip,keepaspectratio]{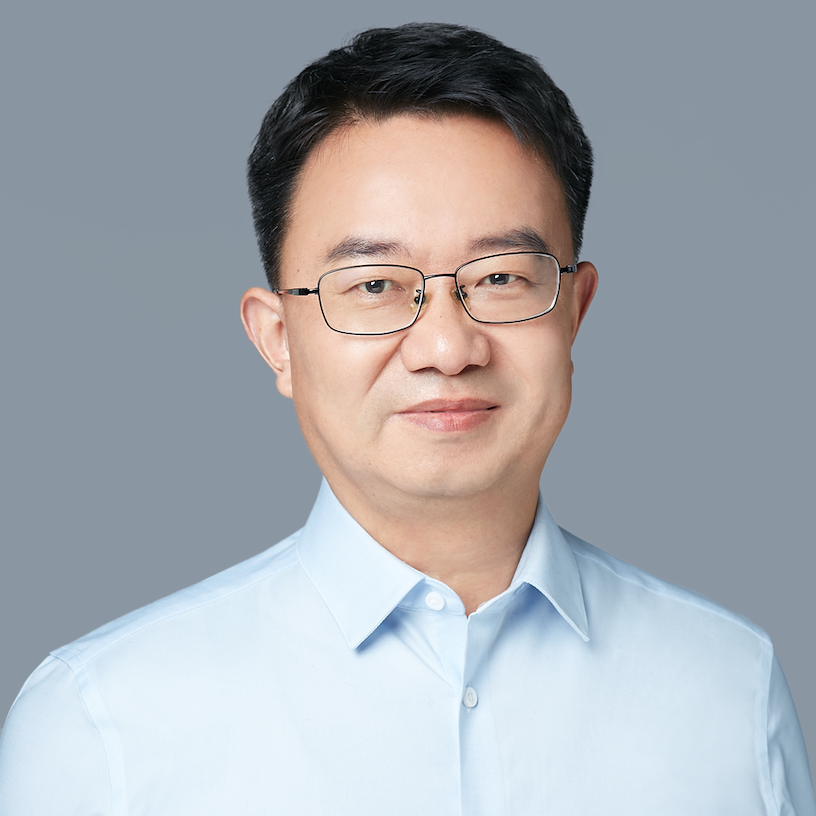}}] {Chuan Shi} (Senior Member, IEEE) received the PhD degree from the ICT of Chinese Academic of Sciences in 2007. He joined the Beijing University ofPosts and Telecommunications as a lecturer in 2007 and is a professor and deputy director of Beijing Key Lab of Intelligent Telecommunications Software and Multimedia at present. His research interests are in data mining, machine learning, and evolutionary computing. He has published more than 40 papers in refereed journals and conferences.
\end{IEEEbiography}

\begin{IEEEbiography} [{\includegraphics[width=1in,height=1.25in,clip,keepaspectratio]{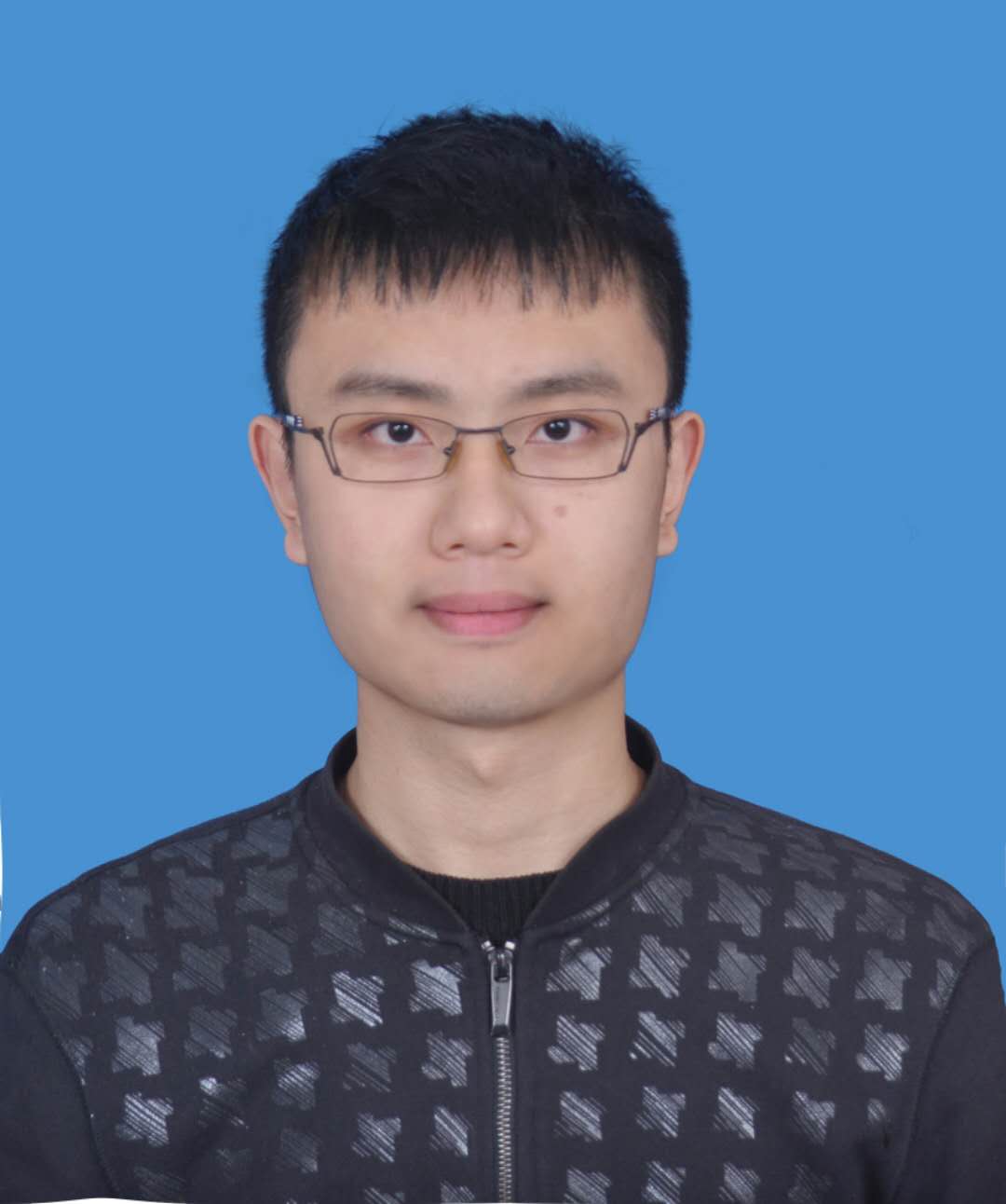}}]{Kaixiang Yang} received the B.S. degree and M.S. degree from the University of Electronic Science and Technology of China and Harbin Institute of Technology, China, in 2012 and 2015, respectively, and the Ph.D. degree from the School of Computer Science and Engineering, South China University of Technology, China, in 2020.

He has been a Research Engineer with the 7th Research Institute, China Electronics Technology Group Corporation, Guangzhou, China, from 2015 to 2017, and has been a postdoctoral researcher with Zhejiang University from 2021 to 2023. He is currently an associate professor with the School of Computer Science and Engineering, South China University of Technology, Guangzhou. His research interests include pattern recognition, machine learning, and industrial data intelligence.
\end{IEEEbiography}

\end{document}